\documentclass[letterpaper,11pt]{article}
\pdfoutput=1
\usepackage{jheppub}
\usepackage{mathtools,amssymb,braket,mathrsfs,tikz,subfig,xspace,xcolor,float,color,siunitx,comment,afterpage,multirow,array,hhline,amsthm,framed}
\usepackage[utf8]{inputenc}
\usepackage[countmax]{subfloat}

\DeclareRobustCommand{\Sec}[1]{Sec.~\ref{#1}}

\DeclareRobustCommand{\App}[1]{App.~\ref{#1}}
\DeclareRobustCommand{\Tab}[1]{Table~\ref{#1}}

\DeclareRobustCommand{\Fig}[1]{Fig.~\ref{#1}}
\DeclareRobustCommand{\Figs}[2]{Figs.~\ref{#1} and \ref{#2}}

\DeclareRobustCommand{\Eq}[1]{Eq.~(\ref{#1})}
\DeclareRobustCommand{\Eqs}[2]{Eqs.~(\ref{#1}) and (\ref{#2})}
\DeclareRobustCommand{\Ref}[1]{Ref.~\cite{#1}}
\DeclareRobustCommand{\Refs}[1]{Refs.~\cite{#1}}
\providecommand{\href}[2]{#2}
\newcommand{\pythia}{\textsc{Pythia}\xspace}
\renewcommand{\O}{\mathcal{O}}

\newtheorem*{conceptdef}{Quark and Gluon Jet Definition (Conceptual)~\cite{Gras:2017jty}}
\newtheorem*{opdef}{Quark and Gluon Jet Definition (Operational)}

\title{An operational definition of quark and gluon jets}

\preprint{MIT--CTP 5042}

\author{Patrick T. Komiske,}
\author{Eric M. Metodiev,}
\author{and Jesse Thaler}

\affiliation{Center for Theoretical Physics, Massachusetts Institute of Technology,\\ 77 Massachusetts Avenue, Cambridge, MA 02139, U.S.A.}

\emailAdd{pkomiske@mit.edu}
\emailAdd{metodiev@mit.edu}
\emailAdd{jthaler@mit.edu}

\abstract{
While ``quark'' and ``gluon'' jets are often treated as separate, well-defined objects in both theoretical and experimental contexts, no precise, practical, and hadron-level definition of jet flavor presently exists.
To remedy this issue, we develop and advocate for a data-driven, operational definition of quark and gluon jets that is readily applicable at colliders.
Rather than specifying a per-jet flavor label, we aggregately define quark and gluon jets at the distribution level in terms of measured hadronic cross sections.
Intuitively, quark and gluon jets emerge as the two maximally separable categories within two jet samples in data.
Benefiting from recent work on data-driven classifiers and topic modeling for jets, we show that the practical tools needed to implement our definition already exist for experimental applications.
As an informative example, we demonstrate the power of our operational definition using $Z$+jet and dijet samples, illustrating that pure quark and gluon distributions and fractions can be successfully extracted in a fully well-defined manner.
}

\begin{document} 
\flushbottom
\maketitle

\section{Introduction}
\label{sec:intro}

Quarks and gluons are fundamental, color-charged particles that are copiously produced at colliders like the Large Hadron Collider (LHC).
Despite their ubiquity, these high-energy quarks and gluons are never observed directly.
Instead, they fragment and hadronize into sprays of color-neutral hadrons, known as \emph{jets}, via quantum chromodynamics (QCD).
As the majority of jets originate from light (up, down, strange) quarks or gluons, a firm understanding of quark and gluon jets is important to many analyses at the LHC.
There has been tremendous recent theoretical and experimental progress in analyzing jets and jet substructure~\cite{Seymour:1991cb,Seymour:1993mx,Butterworth:2002tt,Butterworth:2007ke,Butterworth:2008iy,Abdesselam:2010pt,Altheimer:2012mn,Altheimer:2013yza,Adams:2015hiv,Larkoski:2017jix,Asquith:2018igt}, with a variety of observables~\cite{Berger:2003iw,Almeida:2008yp,Ellis:2010rwa,Thaler:2010tr,Thaler:2011gf,Krohn:2012fg,Larkoski:2013eya,Larkoski:2014uqa,Larkoski:2014pca,Moult:2016cvt,Komiske:2017aww} and algorithms~\cite{Krohn:2009th,Ellis:2009me,Ellis:2009su,Dasgupta:2013ihk,Larkoski:2014wba} developed to expose and probe the underlying physics.
Despite decades of using the notions of ``quark'' and ``gluon'' jets \cite{Nilles:1980ys,Jones:1988ay,Fodor:1989ir,Jones:1990rz,Lonnblad:1990qp,Pumplin:1991kc,Gallicchio:2011xq,Gallicchio:2012ez, Bhattacherjee:2015psa,FerreiradeLima:2016gcz,Bhattacherjee:2016bpy,Komiske:2016rsd,Davighi:2017hok,Cheng:2017rdo,Sakaki:2018opq}, a precise and practical hadron-level definition of jet flavor has not been formulated.

Even setting aside the issue of jet flavor, ambiguity is already present whenever one wants to identify jets in an event~\cite{Salam:2009jx}.
Nonetheless, jets can be made perfectly well-defined: any hadron-level algorithm for finding jets that is infrared and collinear (IRC) safe provides an operational jet definition that can be compared to perturbative predictions.
While different algorithms result in different jets, specifying a jet algorithm allows one to make headway into comparing theoretical calculations and experimental measurements.
Meanwhile, in the case of jet flavor, the lack of a precise, hadron-level definition of ``quark'' and ``gluon'' jets has artificially hindered progress by precluding separate comparisons of quark and gluon jets between theory and experiment.

Typical applications involving ``quark'' and ``gluon'' jets in practice often rely on ill-defined or unphysical parton-level information, such as from the event record of a parton shower event generator.
Progress has been made in providing sharp definitions at the parton-level~\cite{Banfi:2006hf,Buckley:2015gua}, in the context of factorization theorems~\cite{Gallicchio:2011xc,Frye:2016okc,Frye:2016aiz}, and at the conceptual level~\cite{Badger:2016bpw}, but an operational definition, to our knowledge, has never been developed (see \Ref{Gras:2017jty} for a review).
A quark/gluon jet definition%
\footnote{While in some contexts ``jet definition'' means a procedure for finding jets in an event, in this paper we use ``quark/gluon jet definition'' to mean a definition of jet flavor.}
should ideally work at the hadron level, regardless of whether a rigorous factorization theorem exists, and be practically implementable in both theoretical and experimental settings.

In this paper, we develop an operational definition of quark and gluon jets that is formulated solely in terms of experimentally-accessible quantities, does not rely on specific theoretical constructs such as factorization theorems, and can be readily implemented in a realistic context.
Intuitively, we define quark and gluon jets as the ``pure'' categories that emerge from two different jet samples.
Our definition operates at the aggregate level, avoiding altogether the troublesome and potentially impossible notion of a per-jet flavor label in favor of quantifying quark and gluon jets by their distributions.

Specifically, given two jet samples $M_1$ and $M_2$ (e.g.\ $Z+$jet and dijet) in a narrow transverse momentum $(p_T)$ bin, with $M_1$ taken to be more ``quark''-like, and a jet substructure feature space $\mathcal O$, we define quark ($q$) and gluon ($g$) jet distributions in the following way:
\begin{align}
\label{eq:opdefintro}
&p_q(\O)\equiv\frac{p_{M_1}(\O)-\kappa_{12}\,p_{M_2}(\O)}{1-\kappa_{12}},&&p_g(\O)\equiv\frac{p_{M_2}(\O)-\kappa_{21}\,p_{M_1}(\O)}{1-\kappa_{21}},
\end{align}
where $\kappa_{12}$ and $\kappa_{21}$ are known as \emph{reducibility factors} and are directly obtainable from the probability distributions $p_{M_1}(\O)$ and $p_{M_2}(\O)$.
The reducibility factors are defined as:
\begin{align}\label{eq:opkappaintro}
&\kappa_{12} \equiv \min_\mathcal O \frac{p_{M_1}(\mathcal O)}{p_{M_2}(\mathcal O)}, &&\kappa_{21} \equiv \min_\mathcal O \frac{p_{M_2}(\mathcal O)}{p_{M_1}(\mathcal O)}.
\end{align}
The reducibility factors in \Eq{eq:opkappaintro} identify the most $M_1$-like and $M_2$-like regions of the substructure phase space by extremizing the sample likelihood ratio.
We take these phase space regions to \emph{define} what it means to be quark-like and gluon-like.
The subtractions in \Eq{eq:opdefintro} then proceed to ``demix'' the two sample distributions as if they were statistical mixtures.
The quark and gluon distributions are defined solely in terms of hadronic fiducial cross section measurements of the two samples, ensuring that our definition is manifestly fully data-driven and non-circular.
This definition relies on a jet algorithm to define the jets in the jet samples, which also allows for further hadron-level processing, such as jet grooming techniques~\cite{Krohn:2009th,Ellis:2009me,Ellis:2009su,Dasgupta:2013ihk,Larkoski:2014wba}, to be folded directly into the quark/gluon jet definition.

One main goal of this paper is to argue that our operational definition, combined with existing tools, provides a way to obtain information about the likelihood, quark fractions, and quark and gluon distributions in a fully data-driven way, without reference to unphysical notions such as generator labels.
The concepts appearing in our definition are directly related to methods already in use in experimental quark/gluon jet analysis efforts~\cite{CMS-PAS-JME-13-002,Aad:2014gea,Aad:2016oit,CMS-DP-2016-070,ATL-PHYS-PUB-2017-009,Sirunyan:2018asm}.
Quark-gluon likelihood ratios, obtained from parton shower generators, have been implemented by both ATLAS and CMS as optimal discriminants in low-dimensional feature spaces.
Quark fractions, obtained from event generators, for several jet samples have successfully allowed for separate determination of quark and gluon jet properties by solving linear equations.
These analyses already use a statistical-mixture picture of quark and gluon jets, which is a direct consequence of our definition.


Many physics analyses at the LHC would benefit from a clear definition of quark and gluon jets that allows for unambiguous extraction of separate quark and gluon jet distributions and fractions.
Fully data-driven quark/gluon jet taggers have the potential to increase the sensitivity of a variety of new physics searches~\cite{FerreiradeLima:2016gcz,Bhattacherjee:2016bpy}, and related ideas have been developed for model-independent searches for new physics~\cite{Collins:2018epr}.
Experimentally measuring separate quark and gluon distributions of jet observables would significantly improve attempts to extract the strong coupling constant from jet substructure~\cite{Bendavid:2018nar} and to constrain parton shower event generators~\cite{Reichelt:2017hts,Gras:2017jty}.
Extracting data-driven fractions of quark and gluon jets could improve the determination of parton distribution functions and allow for separate measurement of quark and gluon cross sections.
These ideas may also be relevant in the context of heavy ion collisions, where quarks and gluons are expected to be modified differently by the medium and probing the separate modifications to quark and gluon jets would be of significant interest.


We now give a brief summary of the rest of this paper.
In \Sec{sec:def}, we provide a self-contained overview, motivation, and exploration of our quark/gluon jet definition.
We discuss recent work in \Ref{Gras:2017jty} that developed a ``conceptual'' definition of quark/gluon jets, falling short of providing a full definition that can be reliably used in practice, but highlighting the key elements required of a sensible quark/gluon jet definition.
We then develop the intuition and mathematical tools necessary to construct our operational definition, which satisfies the core conceptual principles while being precise and practically implementable.
After stating our operational definition, we examine its physical and statistical properties in detail.
An exploration of the definition in the context of simple jet substructure observables at leading-logarithmic accuracy is left to \App{sec:explore}.

In \Sec{sec:topicwola}, we discuss how our quark/gluon jet definition benefits from, and provides a foundation for, recent work on data-driven machine learning for jet physics.
The classification without labels (CWoLa) paradigm~\cite{Metodiev:2017vrx} for training classifiers on mixed samples can be used to approximate the mixed-sample likelihood ratio, a key part of implementing our definition.
The jet topics framework~\cite{Metodiev:2018ftz} extracts underlying mutually irreducible distributions from mixture histograms, yielding a practical method to obtain the reducibility factors in \Eq{eq:opkappaintro}.
Using jet topics with the approximated mixed-sample likelihood ratio, obtained from the data via CWoLa, allows for more robust fraction and distribution extraction.
With quark fractions, obtained from the data via jet topics, CWoLa classifiers can be (self-)calibrated in a fully data-driven way.
More broadly, the assumptions required for CWoLa and jet topics---that QCD jet samples are statistical mixtures of mutually irreducible quark and gluon jets---are satisfied by construction with our definition.

In \Sec{sec:qgex}, we showcase a practical implementation of our definition using jet samples from two different processes: $Z$+jet and dijets.
Using six trained models detailed in \App{sec:train}, we apply the procedure outlined in \Sec{sec:topicwola} to extract quark fractions by combining the CWoLa and jet topics methods, finding more robust performance than when using single jet substructure observables.
With the reducibility factors and quark fractions in hand, we extract separate quark and gluon distributions for a variety of jet substructure observables, even those that do not exhibit mutual irreducibility.
We compare the results of using our data-driven definition of quark and gluon jets with a per-jet \pythia-parton definition, finding qualitative and quantitative agreement between the two.
The potential to self-calibrate CWoLa classifiers is also shown with an explicit example.
While our studies are based on parton-shower samples, all of these analyses can in principle be performed in data with the experimental tools already developed for quark and gluon jet physics at the LHC.

We present our conclusions in \Sec{sec:conc}, discussing potential new applications made feasible by this work.
Possible future developments and extensions are highlighted.
A study of the similarity of parton-labeled quark and gluon jets between different processes is left to \App{sec:sampledependence}.

\section{Defining quark and gluon jets}
\label{sec:def}

\subsection{Review of a conceptual quark/gluon jet definition}
\label{sec:conceptdef}

Due to the complicated radiative showering and fundamentally non-perturbative hadronization that occurs in the course of jets emerging from partons, there is no unambiguous definition of ``quark'' or ``gluon'' jets at the hadron-level.
Despite this challenge, the importance of a clear, well-defined, and practical definition of quark and gluon jets at modern colliders cannot be overstated.
In \Ref{Gras:2017jty}, a significant effort was made to summarize and comment on the concepts of ``quark jet'' and ``gluon jet''.
The authors of \Ref{Gras:2017jty} settled on the following statement as the best way to conceptually define quark jets (and, analogously, gluon jets):
\clearpage
\begin{conceptdef}
A phase space region (as defined by an unambiguous hadronic fiducial cross section measurement) that yields an enriched sample of quarks (as interpreted by some suitable, though fundamentally ambiguous criterion).
\end{conceptdef}

This definition is attractive for numerous reasons.
First, it is explicitly tied to hadronic final states, avoiding dependence, for example, on the unphysical event record of a parton shower generator.
Further, it is specific to the context of a particular measurement and is thus defined regardless of whether the observable and processes in question have rigorous factorization theorems.
Finally, its goal is to tag a region of phase space as quark- or gluon-like rather than to specify a per-jet truth definition of quark and gluon jets.
The main difficulty with this conceptual definition, as noted in \Ref{Gras:2017jty}, is determining the criterion that corresponds to successful quark or gluon jet enrichment.

Despite its attractive qualities, without a practical proposal for implementing this conceptual definition on data, the case studies in \Ref{Gras:2017jty} operationally fell back on less well-defined definitions, such as using initiating parton information from a parton shower generator to tag a quark/gluon jet.
Further, the definition only tags specific regions of phase space as ``quark'' or ``gluon'', such as low or high values of some substructure observable, and provides no framework for discussing jet flavor outside of these regions.
To remedy this issue, we seek to upgrade the conceptual definition to an operational one by giving a concrete, data-driven method for optimally identifying quark- or gluon-enriched regions of phase space and obtaining full quark and gluon jet distributions.

\subsection{Motivating the operational definition}
\label{sec:motivation}

To motivate our definition, suppose that we have two QCD jet samples $M_1$ and $M_2$ in a narrow $p_T$ bin.
One of the mixed samples ($M_1$ without loss of generality) should be ``quark-enriched'' and the other ``gluon-enriched'' relative to each other according to some qualitative criterion.
\Ref{Gras:2017jty} took $M_1$ and $M_2$ to be, respectively, $Z$+jet and dijet samples, a case that we further investigate in \Sec{sec:qgex}.

Assume for now that $M_1$ and $M_2$ are statistical mixtures of quark and gluon jets---an assumption that will \emph{not} be made in our final definition.
Letting the quark fractions of the two mixtures be $f_1$ and $f_2$, the relationship between the distribution of substructure observables in  mixture $M_i$ in terms of the quark and gluon jet distributions is:
\begin{equation}
\label{eq:mixps}
p_{M_i}(\O)=f_i \, p_q(\O)+(1-f_i)\,p_g(\O),
\end{equation}
where the feature space $\O$ is, for our purposes, a set of jet substructure observables taken to be sufficiently rich to encode all relevant information about jet flavor.

\begin{figure}[t]
\centering
\includegraphics[scale=0.9]{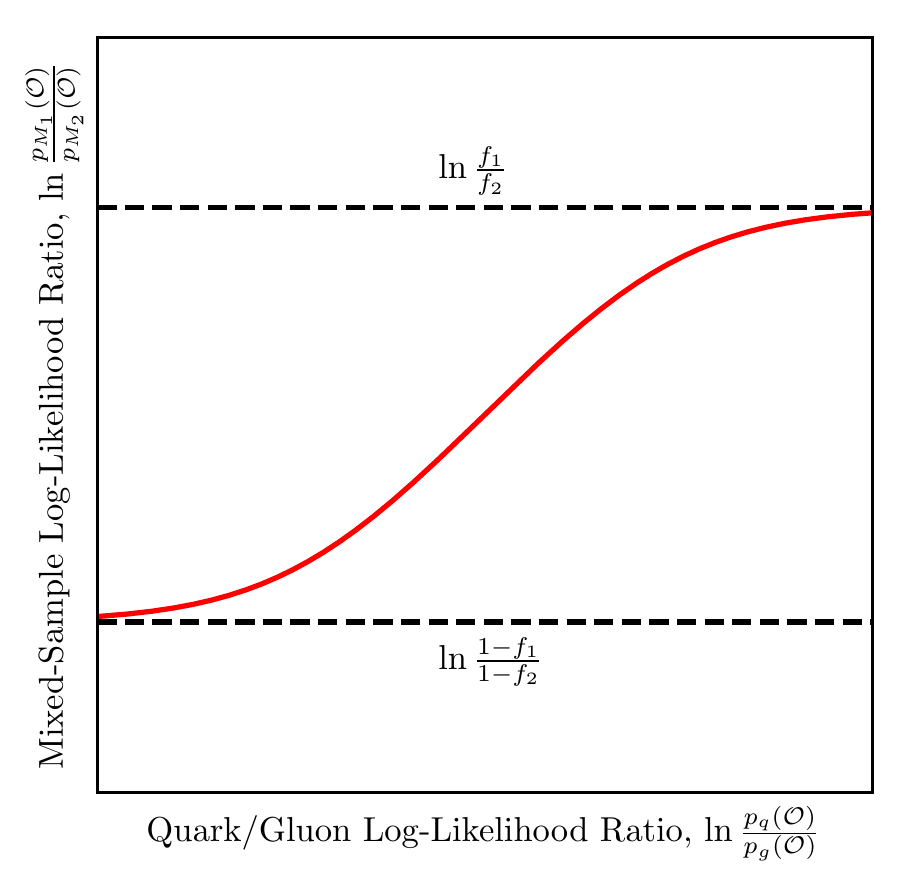}
\caption{
The monotonic relationship between the mixed-sample log-likelihood ratio and the quark-gluon log-likelihood ratio from \Eq{eq:mixnp} for  illustrative fraction values.
The relationship between the maximum and minimum values of the mixed-sample and quark/gluon log-likelihoods from \Eq{eq:kfs} is visually evident in that the red curve horizontally asymptotes to the two black dashed curves.
The plots are shown in terms of the logarithms of the likelihood ratios so that exchanging $M_1\leftrightarrow M_2$ or $q\leftrightarrow g$ simply corresponds to a reflection of the curve.} 
\label{fig:mixloglikes}
\end{figure}

Following the outline of the Conceptual Definition, we consider classification of quark and gluon jets and examine the relationship of this task with classification of one mixture from the other.
By the Neyman-Pearson lemma~\cite{NPlemma}, an optimal classifier for discriminating two classes is their likelihood ratio (or any monotonically-related quantity).
In the case of quark and gluon jets, the likelihood ratio is:
\begin{equation}
\label{eq:np}
L_{q/g}(\O)\equiv\frac{p_q(\O)}{p_g(\O)},
\end{equation}
and, similarly, the optimal classifier for discriminating between $M_1$ and $M_2$ is:
\begin{equation}
\label{eq:mixnp}
L_{M_1/M_2}(\O)\equiv\frac{p_{M_1}(\O)}{p_{M_2}(\O)}=\frac{f_1 \, L_{q/g}(\O)+(1-f_1)}{f_2 \, L_{q/g}(\O)+(1-f_2)}.
\end{equation}
It is easily verified that the mixed-sample likelihood ratio in \Eq{eq:mixnp} is a monotonic function of the quark-gluon likelihood ratio in \Eq{eq:np} as long as $f_1\neq f_2$ (see \Refs{blanchard2016classification,Metodiev:2017vrx}).
The relationship between the mixed-sample likelihood ratio and the quark-gluon likelihood ratio of \Eq{eq:mixnp} is depicted in \Fig{fig:mixloglikes}.
This cleanly demonstrates that the optimal mixed-sample classifier is also the optimal quark-gluon classifier.

Supposing that we can approximate the mixture likelihood ratio sufficiently well, we have distilled the (potentially huge) substructure feature space to a single number which is provably optimal for identifying quark- and gluon-enriched phase space regions.
However, we still lack a procedure for actually identifying the enriched regions; we solely know that they are given by some cut on $L_{q/g}(\O)$, or equivalently a cut on $L_{M_1/M_2}(\O)$.
The key insight for moving closer toward an operational definition is that $L_{q/g}(\O)$, being the optimal discriminant of quark and gluon jets, can be immediately used to identify the most quark-enriched (gluon-enriched) regions as those where $L_{q/g}(\O)$ is at its maximum (minimum). 
In the case that we can find regions of phase space $\mathcal O_q$ and $\mathcal O_g$ where quark and gluon jets respectively are pure, we have that $L_{q/g}(\O_g)=0$ and $L_{g/q}(\O_q)=0$ and we say that the quark and gluon categories are \emph{mutually irreducible} (see \Ref{blanchard2016classification,Metodiev:2018ftz}).

The extrema of the quark/gluon likelihood ratio $L_{q/g}$, corresponding to the enriched regions of phase space, are naturally related to the extrema of the mixture likelihood ratio $L_{M_1/M_2}$.
To this end, it is helpful to define the \emph{reducibility factor} between distributions $A$ and $B$, $\kappa_{AB}$, as:
\begin{equation}
\label{eq:kappa}
\kappa_{AB}\equiv\min_{\O}\frac{p_A(\O)}{p_B(\O)},
\end{equation}
which is the minimum (or more precisely, the infimum) of the likelihood ratio of $A$ and $B$.
Supposing that quarks and gluons are mutually irreducible in the feature space $\O$, the reducibility factors of quark jets to gluon jets (and vice versa) vanish:
\begin{equation}
\label{eq:qgks}
\textbf{Quark and Gluon Jet Mutual Irreducibility}:\quad\quad\kappa_{qg}=0,\quad\quad\kappa_{gq}=0.
\end{equation}

We now show how, assuming quark/gluon mutual irreducibility, the mixture reducibility factors can be related to mixture fractions.
The reducibility factors of the mixed samples can be written down by treating them as mixtures of quarks and gluons as in \Eq{eq:mixps}:
\begin{equation}
\label{eq:mixkappas}
\kappa_{M_iM_j}=\min_{\O} L_{M_i/M_j}(\O)=\min_{\O}\frac{f_i \, L_{q/g}(\O)+(1-f_i)}{f_j \, L_{q/g}(\O)+(1-f_j)}.
\end{equation}
Using our assumptions that $M_1$ is quark-enriched relative to $M_2$, we can write \Eq{eq:mixkappas} as a relation between the mixed-sample reducibility factors and the quark/gluon reducibility factors:
\begin{equation}
\label{eq:mixqgkappas}
\kappa_{M_1M_2}=\frac{f_1\,\kappa_{qg}+(1-f_1)}{f_2\,\kappa_{qg}+(1-f_2)},\quad\quad \kappa_{M_2M_1}=\frac{f_2+(1-f_2)\, \kappa_{gq}}{f_1+(1-f_1)\, \kappa_{gq}},
\end{equation}
where the monotonicity of $L_{M_i/M_j}(\O)$ with $L_{q/g}(\O)$ has been used to push the minimum operation onto the quark-gluon likelihood ratio in \Eq{eq:mixkappas}.
If quarks and gluons are mutually irreducible, we can plug \Eq{eq:qgks} into \Eq{eq:mixqgkappas} to find the reducibility factors of the mixtures:\footnote{An analogous analysis carries through even if non-zero reducibility factors $\kappa_{qg}$ and $\kappa_{gq}$ are specified.}
\begin{equation}
\label{eq:kfs}
\kappa_{12}\equiv\kappa_{M_1M_2}=\frac{1-f_1}{1-f_2},\quad\quad\kappa_{21}\equiv\kappa_{M_2M_1}=\frac{f_2}{f_1}.
\end{equation}
\Fig{fig:mixloglikes} demonstrates that \Eq{eq:mixkappas} defines the asymptotic behavior of the mixed-sample log-likelihood ratio.

Combining the reducibility factors of \Eq{eq:kfs} with the mixture relationship of \Eq{eq:mixps}, we can solve for the underlying quark and gluon jet distributions solely in terms of the well-defined mixture distributions $p_{M_i}(\mathcal O)$ and mixture reducibility factors $\kappa_{ij}$:
\begin{equation}
\label{eq:pqgks}
p_q(\O)=\frac{p_{M_1}(\O)-\kappa_{12}\,p_{M_2}(\O)}{1-\kappa_{12}},\quad\quad p_g(\O)=\frac{p_{M_2}(\O)-\kappa_{21}\,p_{M_1}(\O)}{1-\kappa_{21}}.
\end{equation}
Remarkably, \Eq{eq:pqgks} exposes the underlying quark and gluon jet distributions in terms of experimentally well-defined quantities such as the distribution of jets in mixed samples and their reducibility factors.
Notice also that the quark and gluon distributions each depend on only one of the two mixed-sample reducibility factors. 
Thus, even if only one reducibility factor can be reliably extracted, the corresponding quark or gluon jet distribution can nevertheless be obtained.

Here, we have made several simplifying assumptions, namely that quark and gluon jets can be made well-defined, that $M_1$ and $M_2$ are statistical mixtures of quark and gluon jets, and that quark and gluon jets are mutually irreducible in the feature space $\O$.
\Eq{eq:pqgks} then followed as a consequence, demonstrating that, under these assumptions, it is possible to get access to pure quark and gluon distributions.
What if, on the contrary, we do not make these assumptions, while also requiring that our definition of quark and gluon jets not be circular?
We now proceed to thoroughly explore this idea.

\subsection{An operational definition of quark and gluon jets}
\label{sec:opdef}

We now provide our \emph{operational definition} of quark and gluon jets that builds upon the Conceptual Definition in \Sec{sec:conceptdef} but can be used for practical applications at the LHC and future colliders.
We begin by stating the definition in terms of the notation developed in \Sec{sec:motivation}, and then we proceed to a detailed discussion of its features.

In the absence of any certainty about the underlying structure of samples $M_1$ and $M_2$, we choose to start at the end of \Sec{sec:motivation}, letting \Eq{eq:pqgks} provide a fully-operational definition of quark and gluon jets in terms of experimentally well-defined quantities:
\begin{framed}
\begin{opdef}
Given two samples $M_1$ and $M_2$ of QCD jets at a fixed $p_T$ obtained by a suitable jet-finding procedure, taking $M_1$ to be ``quark-enriched'' compared to $M_2$, and a jet substructure feature space $\O$, the quark and gluon jet distributions are defined to be:
\begin{align}
\label{eq:opdef}
p_q(\O)\equiv\frac{p_{M_1}(\O)-\kappa_{12}\,p_{M_2}(\O)}{1-\kappa_{12}},&&&p_g(\O)\equiv\frac{p_{M_2}(\O)-\kappa_{21}\,p_{M_1}(\O)}{1-\kappa_{21}},
\end{align}
where $\kappa_{12}$, $\kappa_{21}$, $p_{M_1}(\O)$, and $p_{M_2}(\O)$ are directly obtainable from $M_1$ and $M_2$.
\end{opdef}
\end{framed}

There are two immediate points to note about the Operational Definition.
First, it does not attempt to define quark and gluon jets at the level of individual jets, but rather it defines them in aggregate as two well-defined probability distributions.
This is in keeping with the spirit of the Conceptual Definition in \Sec{sec:conceptdef}, which sought to identify enriched regions of phase space rather than to determine a per-jet truth label.
It is also in concert with the basic construction of quantum field theory, which only provides theoretical access to distributional quantities such as cross sections rather than making predictions for individual events.\footnote{Note that (non-deterministic) per-jet labels can be obtained from this definition if needed. For a jet with observable value $O$, one can assign it a ``quark'' label with probability $f\, p_q(O)/(f\, p_q(O) + (1-f)\, p_g(O))$ by using the extracted quark and gluon distributions, $p_q$ and $p_g$, and extracted quark fraction $f$ of the sample. These labels are universal if the observable is monotonically related to the likelihood ratio.}

Second, the Operational Definition does not rely on assumptions of mutual irreducibility of quarks and gluons or the factorization of jet samples as mixtures, instead turning them into derived properties of the definition, as we show below.
In the limit where factorization holds and quarks and gluons are mutually irreducible in the feature space $\mathcal O$, the Operational Definition returns precisely the quark and gluon jets which make sense in that context.
Outside of these potentially-restrictive limits, the definition nonetheless returns two well-defined categories which can be fairly called quark and gluon jets.
The Operational Definition essentially takes the vague notion of ``quark-like'' from the Conceptual Definition and injects mathematical substance by specifying how to extract the quark and gluon distributions.

With the Operational Definition in hand, we now turn the reasoning of \Sec{sec:motivation} on its head to \emph{derive} the mutual irreducibility of quarks and gluons and the mixture nature of the two jet samples $M_1$ and $M_2$.
Using the quark/gluon jet definition in \Eq{eq:opdef}, we can write down the quark/gluon reducibility factors as:
\begin{equation}
\label{eq:derivekqgs}
\kappa_{qg}=\min_{\O}L_{q/g}(\O)=\min_{\O}\frac{(1-\kappa_{21})(L_{M_1/M_2}(\O)-\kappa_{12})}{(1-\kappa_{12})(1-\kappa_{21}L_{M_1/M_2}(\O))}=0,
\end{equation}
where we have used the monotonicity of $L_{q/g}(\mathcal O)$ in $L_{M_1/M_2}(\O)$ and the definition of $\kappa_{12}$ to see that the numerator vanishes while the denominator is non-zero.
An analogous calculation shows that $\kappa_{gq}=0$, and therefore that the distributions of quark and gluon jets as defined by the Operational Definition are always mutually irreducible.

Next, we demonstrate that $M_1$ and $M_2$ are mixtures of the defined quark and gluon jet distributions.
Solving \Eq{eq:opdef} for the distributions of $M_1$ and $M_2$ in terms of the quark/gluon distributions yields:
\begin{align}
\label{eq:solve4m1}
&p_{M_1}(\O)=f_1\, p_q(\O)+(1-f_1)\, p_g(\O),& f_1&\equiv\frac{1-\kappa_{12}}{1-\kappa_{12}\kappa_{21}},\\
&p_{M_2}(\O)=f_2 \, p_q(\O)+(1-f_2)\, p_g(\O),& f_2&\equiv\frac{\kappa_{21}(1-\kappa_{12})}{1-\kappa_{12}\kappa_{21}},
\label{eq:solve4m2}
\end{align}
where we have introduced two numbers $f_1$ and $f_2$ such that $f_1,f_2\in[0,1]$.
We see from \Eqs{eq:solve4m1}{eq:solve4m2} that under the Operational Definition, $M_1$ and $M_2$ have the interpretation of being statistical mixtures of quark and gluon jets where the quark fractions of each sample are $f_1$ and $f_2$, respectively.
Note that while this was entirely anticipated, given the motivation provided in \Sec{sec:motivation}, the Operational Definition manages to avoid the circular reasoning of that section, where a well-defined notion of quark and gluon jets and the statistical-mixture nature of $M_1$ and $M_2$ were assumed to exist before we were able to specify a rigorous procedure to determine them.

There are several additional properties of the Operational Definition worth noting.
First, any additional preprocessing of the jets in $M_1$ and $M_2$ which is operationally defined at the hadron level, such as jet grooming, can be folded into the jet-finding procedure and thus incorporated directly into our definition.
Second, which of $M_1$ or $M_2$ is more ``quark-enriched'' only serves to label which of the resulting distributions is ``quark'' and which is ``gluon'' and does not change the distributions which are produced by this definition.
Finally, while \Eq{eq:opdef} implies the vanishing of the quark/gluon reducibility factors, if a different, non-zero quark/gluon reducibility factor is desired a priori, then the definition may be suitably modified to accommodate those non-zero values.
Thus, the assertion of quark-gluon mutual irreducibility, which is supported by evidence from case studies, can be relaxed to any specified quark/gluon reducibility factors which may then be thought of as inputs to the definition.

In \Sec{sec:topicwola}, we connect the Operational Definition to machinery that has already been developed in the jet substructure and statistical literature, finding that the tools needed to implement the Operational Definition, true to the name, are readily available.
In \App{sec:explore}, we gain some additional insight into the Operational Definition by theoretically exploring it with simple jet substructure observables in a tractable limit of perturbative QCD.

\section{Data-driven jet taggers and topics}
\label{sec:topicwola}

In this section, we connect our Operational Definition of quark and gluon jets to recent developments at the intersection of jet physics and statistical methods, particularly the data-driven paradigms of CWoLa~\cite{Metodiev:2017vrx} and jet topics~\cite{Metodiev:2018ftz}.
CWoLa provides a method to approximate the quark/gluon likelihood ratio by distilling the available information in a huge feature space of jet substructure observables~\cite{Metodiev:2017vrx, Cohen:2017exh, Komiske:2018oaa}.
The jet topics method was introduced and developed in \Ref{Metodiev:2018ftz}, where it was shown that statistical methods could be used to ``disentangle'' quark and gluon jets from mixtures.
We will show how these methods can be combined to form a concrete implementation of the Operational Definition.

\subsection{Classification without labels: Training classifiers on collider data}
\label{sec:cwola}

Recently, there has been an effort to train physics classifiers directly on data despite the lack of labeled truth information, going under the broad term of \emph{weak supervision}.
\Ref{Dery:2017fap} was the first to apply weak supervision methods in a particle physics context, showing that given mixed samples with known signal fractions, a quark/gluon classifier on a few high-level inputs could be trained without access to per-jet truth labels, a paradigm termed learning from label proportions (LLP).
\Ref{Metodiev:2017vrx} developed CWoLa as a method to train a jet classifier via weak supervision on a few generalized angularities~\cite{Berger:2003iw,Almeida:2008yp,Ellis:2010rwa,Larkoski:2014uqa,Larkoski:2014pca}, where signal fractions do not need to be known in order to train the classifier.
\Ref{Komiske:2018oaa} investigated both CWoLa and LLP in the context of high-dimensional, modern machine learning methods, finding that while both methods were performant, CWoLa generalized better and more simply to complex models.
CWoLa has since given rise to new techniques to search for signals of new physics in model-independent ways~\cite{Collins:2018epr}.
These methods are an important step towards making classification at colliders fully data-driven.
Here, we review the CWoLa paradigm in preparation for incorporating it as part of the implementation of our Operational Definition.

Conceptually, CWoLa is extremely simple: given two mixtures $M_1$ and $M_2$ of signal (quark) and background (gluon) jets, train a classifier to distinguish jets in $M_1$ from jets in $M_2$.
This procedure has the attractive property of being able to immediately use any model which can be trained with full supervision.
Furthermore, in the limit that $M_1$ and $M_2$ become pure signal and background, CWoLa smoothly approaches full supervision.
With enough statistics, a feature space that captures all relevant information, and a suitable training procedure, a CWoLa classifier should approach the optimal discriminant between the two mixed samples.\footnote{The generalization to learning from multiple mixtures of signal and background is possible as long as each mixture is assigned a label that is (on average) monotonically related to its signal fraction.}
By the Neyman-Pearson lemma~\cite{NPlemma}, the optimal discriminant between two binary classes is the likelihood ratio.
As discussed in \Sec{sec:motivation}, the mixed-sample likelihood ratio is monotonically related to the quark/gluon jet likelihood ratio.
Thus, CWoLa provides a way of approximating the optimal discriminant between quark and gluon jets given access only to mixed samples.

There are potential concerns, though, that one might have regarding CWoLa in particular and weak supervision in general.
Are enough statistics and a rich-enough feature space available?
Do we have a suitable training procedure?
\Refs{Metodiev:2017vrx, Cohen:2017exh, Komiske:2018oaa} address these concerns and demonstrate that CWoLa indeed works in realistic cases.
For example, CWoLa was used in \Ref{Komiske:2018oaa} to obtain a performant quark/gluon jet classifier by discriminating $Z$+jet and dijet samples using jet images and convolutional neural networks.
As described in \App{sec:train}, there are many other jet representations and machine learning models that are suitable to be trained with CWoLa.
Additionally, previous uses of CWoLa have made assumptions about the samples $M_1$ and $M_2$ being mixtures of well-defined quark and gluon jets, without specifying which definition is being used or attempting to quantify what happens if quark and gluon jets are not the same in the two samples (i.e.\ sample dependence).
From the perspective of this work, those concerns are removed by using the Operational Definition, which turns the problem on its head and lets the samples $M_1$ and $M_2$ \emph{define} quark and gluon jets.
The notion of sample dependence manifests in a new way with our Operational Definition, which we discuss more in our conclusions in \Sec{sec:conc}.

\subsection{Jet topics: Extracting categories from collider data}
\label{sec:topics}

Building on a rich analogy between mixed jet samples and textual documents, \Ref{Metodiev:2018ftz} introduced jet topics and demonstrated how topic modeling could be used to obtain quantitative information about the signal and background distributions from the mixed sample distributions.
The present work extends and elaborates on this approach in order to formulate a practical implementation the Operational Definition of quark and gluon jets in \Sec{sec:opdef}.

Given two samples of quark and gluon jets $M_1$ and $M_2$, the jet topics technique seeks to extract two mutually irreducible categories such that the samples are mixtures of these categories.
To the extent that quark and gluon jets are themselves mutually irreducible, they will correspond to the extracted topics.
There are various procedures for extracting the topics from mixed samples.
\Ref{Metodiev:2018ftz} used a method known as ``demixing'' that was developed in \Ref{katz2017decontamination} in order to obtain the topics.
Other procedures (e.g.\ non-negative matrix factorization~\cite{Arora:2012:LTM:2417500.2417847}) that are popular for textual topic modeling could in principle also be used.
Demixing works by searching for ``anchor bins'' in the mixed sample distributions over a feature space $\O$, which are histogram bins for which the likelihood of $M_1$ to $M_2$ is maximized or minimized.

In the language of \Sec{sec:motivation}, demixing returns reducibility factors $\kappa_{12}$ and $\kappa_{21}$.
With the reducibility factors in hand, the fractions of topic $T_1$ in each mixed sample, $f_{T_1}^{(1)}$ and $f_{T_1}^{(2)}$, can be obtained by solving equations analogous to \Eq{eq:kfs}, and the topic distributions $p_{T_1}(\O)$ and $p_{T_2}(\O)$ are given by \Eq{eq:pqgks} where $q$ is replaced by $T_1$ and $g$ by $T_2$:
\begin{align}
\label{eq:topic1}
&p_{T_1}(\O)=\frac{p_{M_1}(\O)-\kappa_{12}\, p_{M_2}(\O)}{1-\kappa_{12}},&f_{T_1}^{(1)}&=\frac{1-\kappa_{12}}{1-\kappa_{12}\kappa_{21}},\\
&p_{T_2}(\O)=\frac{p_{M_2}(\O)-\kappa_{21}\, p_{M_1}(\O)}{1-\kappa_{21}},&f_{T_1}^{(2)}&=\frac{\kappa_{21}(1-\kappa_{12})}{1-\kappa_{12}\kappa_{21}},
\label{eq:topic2}
\end{align}
where we have assumed without loss of generality that $f_{T_1}^{(1)}>f_{T_1}^{(2)}$.

The jet topics method provides a simple example of the fascinating mileage one is able to achieve from the picture of jets as statistical mixtures.
If the signal (quark) and background (gluon) distributions are mutually irreducible, the topic fractions are the signal fractions, $f_S^{(1)}=f_{T_1}^{(1)}$ and $f_S^{(2)}=f_{T_1}^{(2)}$, from which a number of other useful quantities may be computed.
First, consider some observable $O$ that we wish to cut on to make a signal/background classifier. 
For a given threshold $t$, let the fraction of jets in $M_i$ for which $O$ is greater than $t$ be $f_{M_i}(O>t)$.
Let $\varepsilon_{s}(t)$ be the rate that the signal is correctly identified (the true positive rate) and $\varepsilon_{b}(t)$ be the rate that the background is identified as signal (the false positive rate) by the classifier $(O, t)$.
We then have the equations:
\begin{align}
\label{eq:calibrate}
&f_{M_1}(O>t)=f_S^{(1)}\varepsilon_s(t) + (1-f_S^{(1)})\, \varepsilon_b(t),\\
&f_{M_2}(O>t)=f_S^{(2)}\varepsilon_s(t) + (1-f_S^{(2)})\, \varepsilon_b(t),
\end{align}
which can be solved to give signal and background efficiencies at the given threshold:
\begin{align}
\label{eq:eps}
\varepsilon_s(t)&=\frac{f_{M_1}(O>t)(1-f_S^{(2)}) - f_{M_2}(O>t)(1-f_S^{(1)})}{f_S^{(1)} - f_S^{(2)}},\\
\label{eq:epb}
\varepsilon_b(t)&=\frac{f_{M_2}(O>t)f_S^{(1)} - f_{M_1}(O>t)f_S^{(2)}}{f_S^{(1)} - f_S^{(2)}}.
\end{align}
In this way, the extracted fractions can be used to calibrate the classifier.
Additionally, the pure signal and background distributions of any observable can be obtained from the reducibility factors (or equivalently the extracted fractions): simply change the feature space $\O$ in \Eqs{eq:topic1}{eq:topic2} to whatever observable is desired.

There are several issues to address in attempting to use topic modeling for quark and gluon jets.
How do we know that quark and gluon jets are mutually irreducible in our feature space?
In \App{sec:explore}, we show that quark and gluon jets are {\it not} mutually irreducible in the leading-logarithmic limit of individual Casimir-scaling or Poisson-scaling observables, though this calculation strongly suggests that mutual irreducibility could be achieved in a larger feature space.
\Ref{Metodiev:2018ftz} showed that quark and gluon jets appear to be mutually irreducible in practice for the constituent multiplicity observable, but did not offer a way to fold in additional information.
If we attempt to use multiple observables in the topic modeling procedure, how do we deal with the curse of dimensionality that results from attempting to fill multi-dimensional histograms?
As we now discuss, CWoLa can be combined with jet topics to efficiently use arbitrarily large feature spaces to determine the optimal quark and gluon jet topics.

\subsection{Optimal taggers for optimal topics}
\label{sec:optimal}

To summarize, the CWoLa framework allows trained models to approximate a function monotonic to the quark/gluon likelihood ratio, which is the optimal quark/gluon jet classifier.
Further, the jet topics technique allows for signal and background distributions to be extracted from a given low-dimensional feature space.
Here, we demonstrate how CWoLa and jet topics can be combined into a direct implementation of the Operational Definition of quark and gluon jets from \Sec{sec:opdef}.

When viewed as a likelihood-ratio approximator, a CWoLa-trained model can do more than per-jet classification: it is an efficient method for compressing information in a (potentially) huge but sparsely-populated feature space down to the provably optimal single observable for quark/gluon jet separation.
This approach of taking a CWoLa-trained model output as an interesting observable in its own right solves the curse of dimensionality mentioned at the end of \Sec{sec:topics}.
Furthermore, the guarantee of optimality for the likelihood ratio by the Neyman-Pearson lemma carries over to the jet topics context in that the mutual irreducibility of quark and gluon jets is maximized when the optimal discriminant is used. 
In this sense, optimal taggers give rise to optimal topics.

The marriage of CWoLa and jet topics yields more fruit: since the signal fractions extracted by the topics procedure can be used to calibrate a classifier, the requirement that a CWoLa-trained model be calibrated using known signal fractions is removed.
A CWoLa model now has the potential to be \emph{self-calibrating} in the sense that the model is used to extract the signal fractions, and then the fractions are used to calibrate that same model (other models can also be calibrated).
Furthermore, the optimal topic fractions can be used to extract the pure distribution of any desired observable in a straightforward manner.

This combined paradigm provides a new way to use fully data-driven classifiers in high-energy particle physics, namely as optimal observables for topic fraction extraction. 
The fully data-driven aspect of the entire procedure cannot be emphasized enough as application of these methods to data is the ultimate goal.
The black-box nature of complex classifiers becomes less disturbing in this context, since we can think of the role of the classifier as simply to regress onto the likelihood ratio, without much concern as to how this is done.
As with \Ref{Andreassen:2018apy}, understanding of both the inputs and outputs of a machine learning model allows us to be agnostic with respect to the internal details.

Where does the Operational Definition in \Sec{sec:opdef} fit into this picture?
If we adopt the Operational Definition and define quark and gluon jets to be the categories returned by the topic-finding procedure, this addresses the first issue with jet topics referenced at the end of \Sec{sec:topics}, that we do not know the relation between the extracted topics and quark and gluon jets.
Also, since under this definition the samples $M_1$ and $M_2$ are mixtures of exactly the same quark and gluon jets, the sample dependence concerns mentioned at the end of \Sec{sec:cwola} are alleviated.
The optimality guarantee resulting from the Neyman-Pearson lemma and the good practical performance lend support to the Operational Definition being useful both in theory and practice.
It is no coincidence that the Operational Definition, CWoLa, and jet topics share the same property: they work well when notions of sample independence and mutual irreducibility exist, but still return something sensible as the situation is detuned away from this nice limit.

\section{Quark and gluon jets from dijets and $Z$+jet}
\label{sec:qgex}

In this section, we apply the combined paradigm of CWoLa and jet topics to the realistic context of $Z$+jet and dijet samples, obtaining the distributions of quark and gluon jets via the Operational Definition.%
\footnote{We also investigated applying the Operational Definition to CMS jet mass measurements on similar samples~\cite{Chatrchyan:2013vbb}.  In the dijet sample, though, only average jet mass (instead of individual jet mass) is reported.} 

\subsection{Event generation}
\label{sec:eventgen}

We generated events using \pythia 8.230~\cite{Sjostrand:2014zea} with the default tunings and shower parameters at $\sqrt{s}=\SI{14}{TeV}$.
Hadronization and multiple parton interactions (i.e.\ underlying event) were included and a parton-level $p_T$ cut of $\SI{400}{GeV}$ was applied. 
The $Z$+jet sample was obtained using the \texttt{WeakBosonAndParton:qg2gmZq} and \texttt{WeakBosonAndParton:qqbar2gmZg} processes, ignoring the photon contribution and requiring the $Z$ to decay invisibly.
The dijet sample was obtained using the \texttt{HardQCD:all} process, excluding bottom quarks.

Final state, non-neutrino particles were clustered with \textsc{FastJet} 3.3.0~\cite{Cacciari:2011ma} using the anti-$k_T$ algorithm~\cite{Cacciari:2008gp} with a jet radius of $R=0.4$.
All jets were required to have $p_T\in[500,550]$ GeV and rapidity $|y|<2.5$.
The hardest jet for $Z$+jet and the hardest two jets for dijets were considered and kept if they passed the above specified cuts.
The unphysical parton-shower-labeled jet flavor was determined by matching the clustered jet to the \pythia parton(s) by requiring that the jet lie within $2R$ of the parton direction from the hard process.
Events in which none of the jets passed this criteria were not considered.
One million jets passing all cuts were retained for both the dijet and $Z$+jet samples. 
The \pythia-labeled quark fraction was 86.3\% for the $Z$+jet sample and 49.8\% for the dijet sample.

\begin{table}[t]
\centering
\begin{tabular}{|c|l|l|}
\hline
 Symbol & Name &Short Description  \\ \hline \hline
$n_\text{const}$ & Constituent Multiplicity & Number of particles in the jet \\
$n_\text{SD}$ & Soft Drop Multiplicity & Probes number of perturbative emissions \\
$N_{95}$ & Image Activity & Number of pixels containing $95\%$ of jet $p_T$\\
$\tau_2^{(\beta=1)}$ & 2-subjettiness & Probes the two-prong nature of the jet\\
$w$ &  Width & Angularity measuring the girth of the jet\\
$m$ & Jet Mass & Mass of the jet\\
\hline \hline
PFN-ID & Particle Flow Network with ID & Particle three-momentum + ID inputs\\
PFN & Particle Flow Network & Particle three-momentum inputs \\
EFN & Energy Flow Network & Using only IRC-safe information\\
EFPs & Energy Flow Polynomials & Linear classification with EFPs\\
CNN & Convolutional Neural Network & Trained on $33\times 33$ 2-channel jet images\\
DNN & Deep Neural Network & Trained on an $N$-subjettiness basis \\
\hline
\end{tabular}
\caption{The individual jet substructure observables (top) and machine learning models (bottom) considered in this study, along with their corresponding symbols and short descriptions.
A full discussion of the observables and models is given in \App{sec:train}.}
\label{tab:obsandmodels}
\end{table}

\subsection{Extracting reducibility factors and fractions}
\label{sec:extract}

For the jet substructure feature space $\mathcal O$, we consider a variety of individual jet substructure observables and trained models.
In \Tab{tab:obsandmodels}, we summarize the observables and models used for our study.
Details of the observable computation, model training, and model architectures are given in \App{sec:train}.

For each of the observables and trained models, we proceed to extract the topic fractions from the $Z$+jet and dijet samples.
We implement a version of the demixing procedure used in \Ref{Metodiev:2018ftz} and described in \Ref{katz2017decontamination}.
Below, we describe the practical procedure used for the studies in this section, including the determination of uncertainties.
Here, we let $O$ indicate either a single observable or the output of a trained model.

\begin{enumerate}
\item {\bf Histograms}: The histograms for $p_\text{$Z+$jet}(O)$ and $p_\text{dijets}(O)$ are computed for a specified binning. Statistical uncertainties are taken to be $\sqrt{N_\text{$Z+$jet}}$ and $\sqrt{N_\text{dijets}}$ coming from one-sigma count uncertainties within each bin.\footnote{These uncertainties, and those derived from them, should only be used to give a sense of scale on the plots. Implementing the Operational Definition in LHC data will require careful consideration of other sources of statistical and systematic uncertainties. For instance, using unfolded distributions may mitigate artificial differences in the samples due to detector effects.}
\item {\bf Likelihood Ratios}: The mixed-sample log-likelihood ratio $\ln p_\text{dijets}(O)/p_\text{$Z+$jet}(O)$ is calculated. The statistical uncertainty is estimated from uncertainty propagation per bin to be:
\begin{equation}
\sigma_{\ln p_\text{dijets}/p_\text{$Z+$jet}} = \sqrt{\frac{1}{N_\text{dijets}} + \frac{1}{N_\text{$Z+$jet}}}.
\end{equation}
\item {\bf Anchor Bins}: Noisy, low-statistics bins are neglected by only considering bins with more than 50 events in each sample. The upper (lower) anchor bin is obtained by finding the maximum (minimum) bin for the log-likelihood ratio minus (plus) its uncertainty.
\item {\bf Reducibility Factors}: The lower (upper) reducibility factor $\kappa_{21}$ ($\kappa_{12}$) is obtained by exponentiating (minus) the log-likelihood ratio evaluated at the lower (upper) anchor bin. Uncertainties on the reducibility factors are obtained by standard uncertainty propagation.
\item {\bf Topics}: The jet topics are obtained from the reducibility factors $\kappa_{12}$ and $\kappa_{21}$ according to the definition in \Eq{eq:opdef}, with uncertainties propagated from the reducibility factors.
\item {\bf Fractions}: Topic fractions are obtained from the reducibility factors $\kappa_{12}$ and $\kappa_{21}$ according to \Eqs{eq:solve4m1}{eq:solve4m2}, with uncertainties propagated from the reducibility factors.
In this study, the topic fraction always corresponds to the quark fraction.
\end{enumerate}

While we use the concrete method above to showcase the viability of our method, there may of course be alternative ways to obtain the anchor bins and reducibility factors.
For instance, it may be interesting to a pursue a binning-free method, where a cumulative density function is used instead of a binned histogram.
Similarly, there may be more suitable ways to ignore low-statistics phase space regions and determine anchor bins.
We leave detailed optimizations of the method for future developments.

\begin{figure}[t]
\centering
\subfloat[]{\includegraphics[scale=.735]{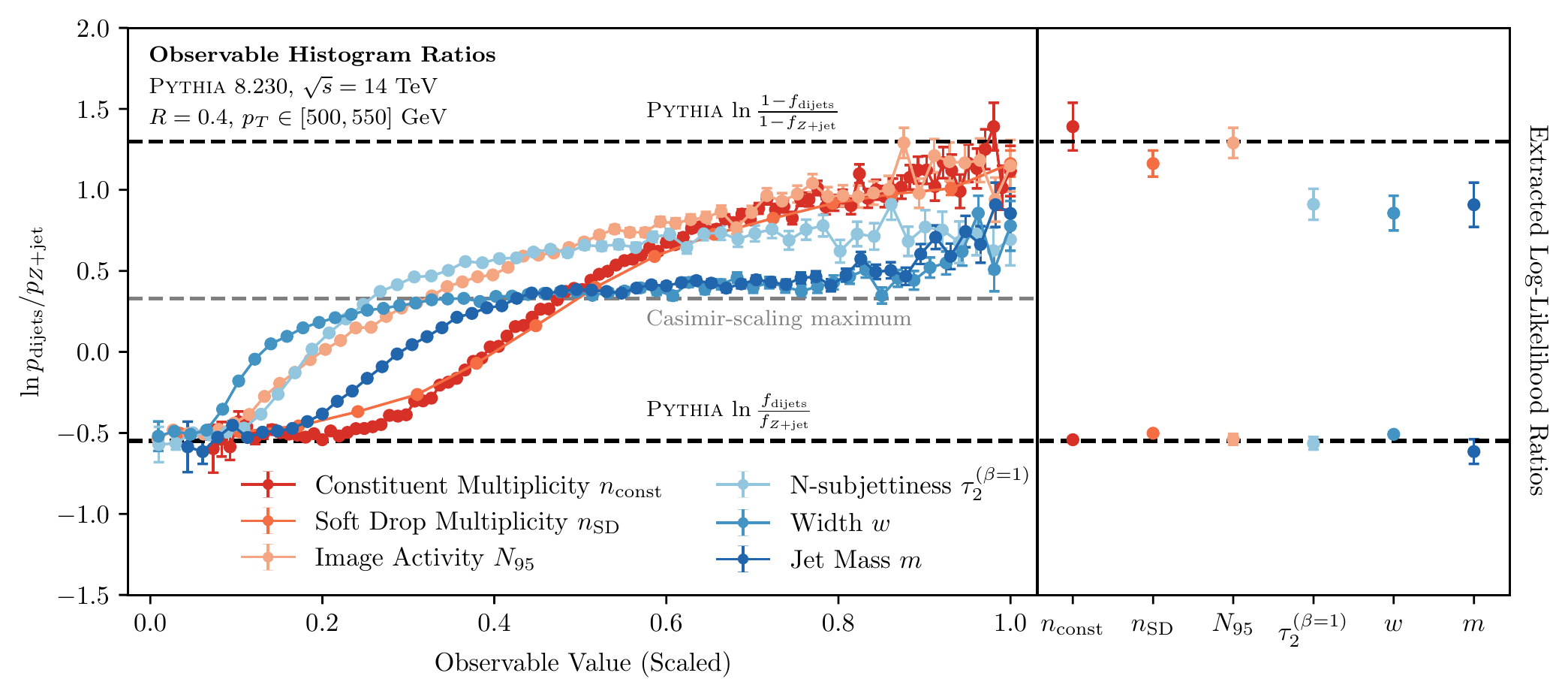}\label{fig:histratios-obs}}

\subfloat[]{\includegraphics[scale=.735]{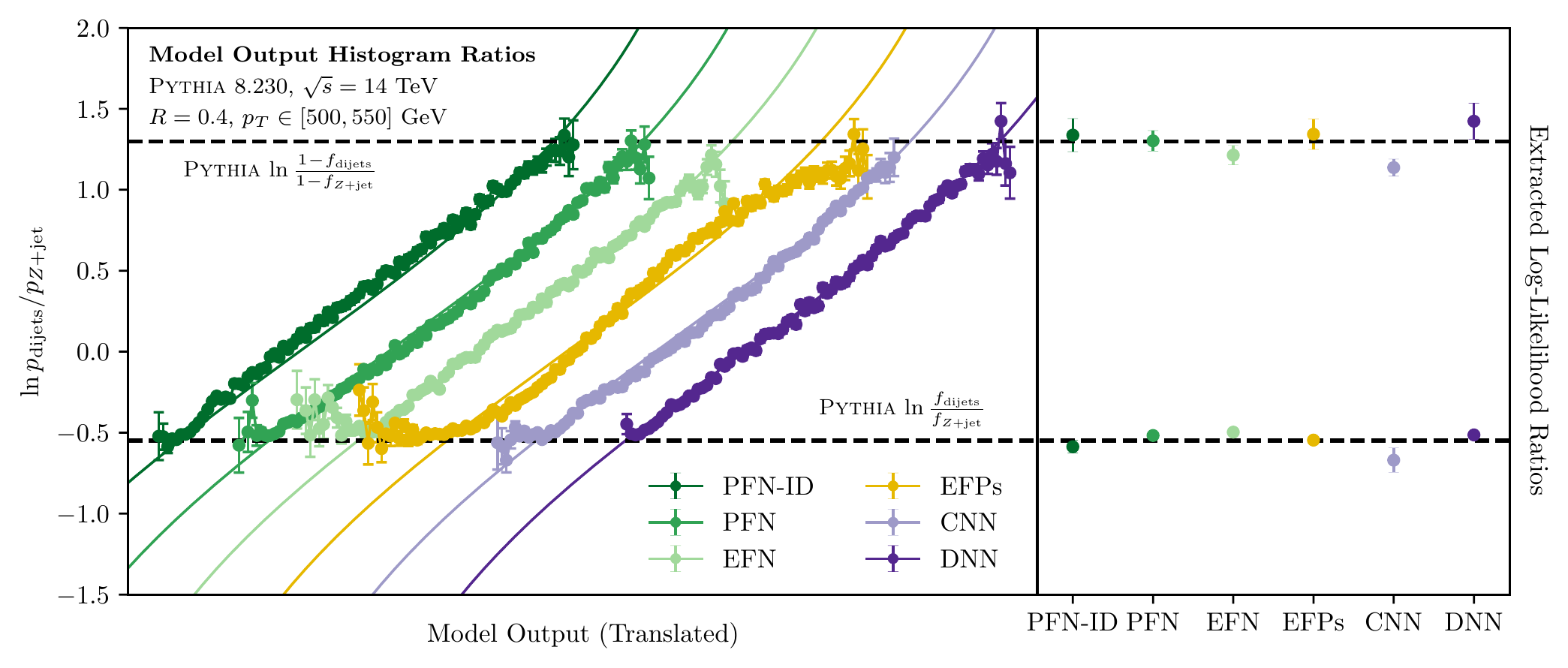}\label{fig:histratios-model}}
\caption{
Mixture log-likelihood ratios and their extrema for (a) individual jet substructure observables and (b) trained models, the latter of which have been translated along the horizontal axis for clarity. 
The black dashed lines indicate the maximum and minimum of the mixture likelihood ratio determined using the \pythia fractions.
The gray dashed line in the observable plot indicates the upper bound obtained for a Casimir-scaling observable from \App{sec:explore}; as expected, jet mass and width approach and remain near the gray line for much of their domain.
While all individual observables asymptote well to the lower black line, only the count observables ($n_{\rm const}$, $n_{\rm SD}$, $N_{95}$) come close to the upper black line, indicating that gluons are more irreducible than quarks.
By contrast, the minimum and maximum for each trained model appear to achieve extremal values close to the black limits.
The solid colored lines in the lower plot indicate the behavior of the optimal classifier, closely related to \Fig{fig:mixloglikes}.
}
\label{fig:histratios}
\end{figure}

In \Fig{fig:histratios}, we show the mixed-sample log-likelihood ratios $\ln p_\text{dijets}(O)/p_\text{$Z+$jet}(O)$ for various jet substructure observables and model outputs.
Overall, we see excellent confirmation that the mixed-sample log likelihood is bounded between the predicted extrema according to the \pythia fractions.
To extract these fractions in a data-driven way, we must of course obtain these extrema from the measured log-likelihood ratios.
Using the procedure outlined above, the resulting anchor bins are shown in the right-most portion of \Fig{fig:histratios}.
Interestingly and satisfyingly, many of the individual observables and essentially all of the models extract extrema consistent with the \pythia fractions.
It is important to note, though, that the \pythia fractions are not fully well-defined hadron-level concepts and are shown solely to provide a conceptual and semi-quantitative guideline for the performance of the method.

\afterpage{\clearpage}

For the substructure observables in \Fig{fig:histratios-obs}, it is evident that the count observables of constituent multiplicity, soft drop multiplicity, and image activity come closest to saturating both the upper and lower bounds.
For mass and width, a clear plateau is exhibited close to the leading logarithmic expectation for Casimir-scaling observables (see \App{sec:explore}).
This difference is reflected in the fact that the count observables extract extrema of the log-likelihood ratio consistent with the \pythia fractions, while the remaining observables systematically underestimate the upper bound.
One feature worth noting is that the lower bound is accurately extracted by every observable; it is the upper bound that is more difficult to saturate with a generic observable.
This indicates that gluon jets are evidently more irreducible than quark jets, and therefore that gluon jet distributions are easier to extract.

For the trained model outputs in \Fig{fig:histratios-model}, we see that the mixed-sample log-likelihood ratios are clearly bounded as expected and agree with the prediction for a well-trained classifier.
The slight deviations from the solid curve in the case of the EFPs arise from the fact that they are trained using Fisher's Linear Discriminant, which optimizes a different objective function, but nonetheless the EFPs exhibit qualitatively similar behavior to the other classifiers.
Compared to the individual substructure observables, the models more robustly saturate the upper and lower bounds of the log-likelihood ratio and demonstrate less sensitivity to changes in the binning of the histograms.
The extracted extrema of the log-likelihood ratio based on the trained models (with the exception of the CNN) are all consistent with one another as well as with the \pythia fractions.
This agreement, present in the variety of different models which process information in very different ways, indicates that there is indeed a robust sense in which ``quark'' and ``gluon'', as qualitatively described by the parton-matched labels, are latent within the mixed samples.

\begin{figure}[t]
\centering
\subfloat[]{\includegraphics[scale=.755]{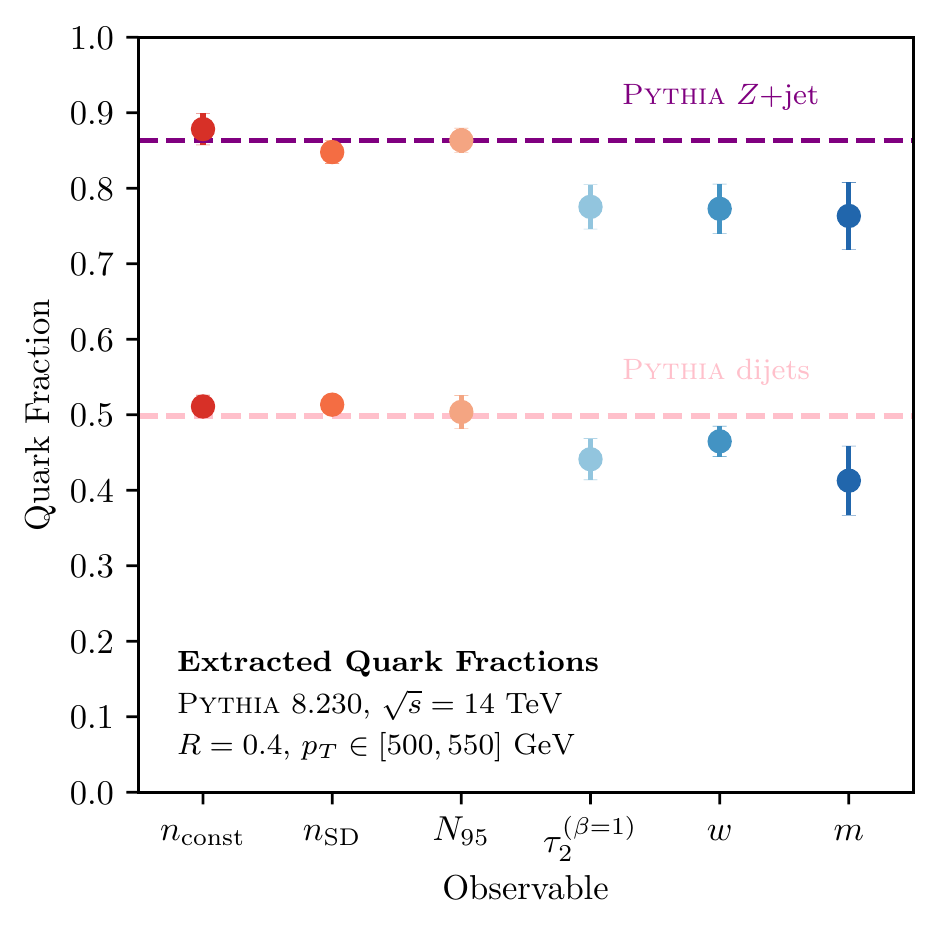}\label{fig:fracs-obs}}
\subfloat[]{\includegraphics[scale=.755]{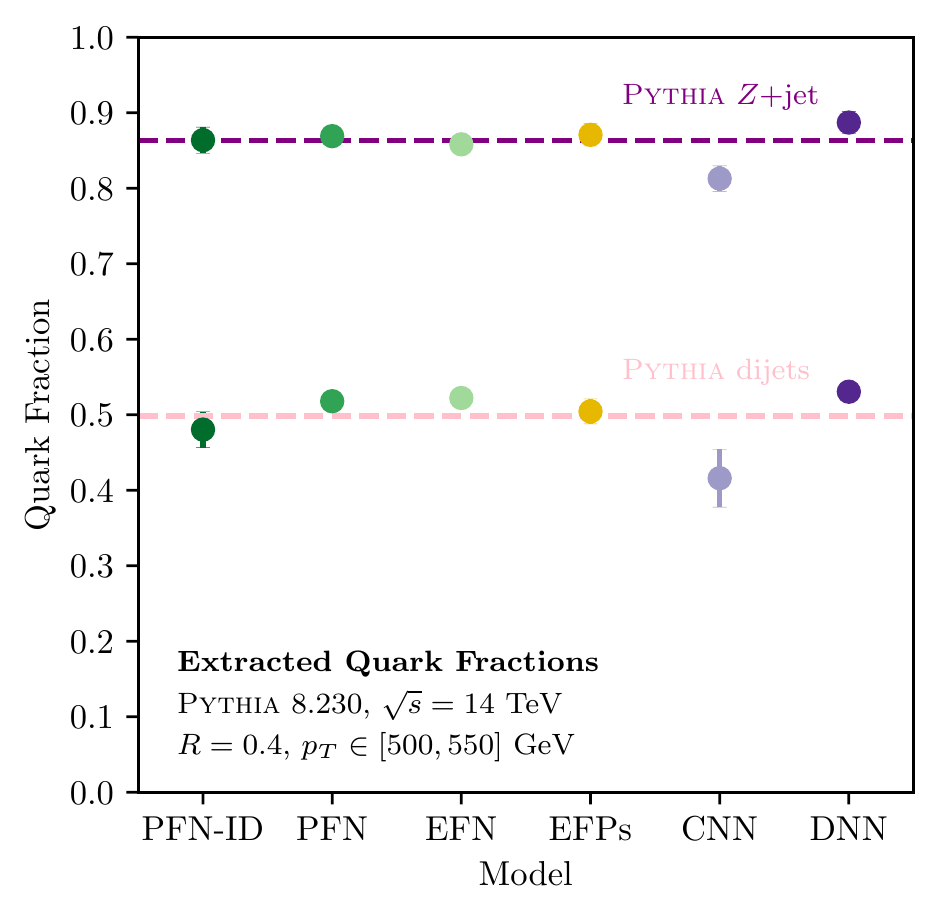}\label{fig:fracs-model}}
\caption{
Extracted quark fraction values for the (a) individual observables and (b) trained models as calculated using the log-likelihood extrema of \Fig{fig:histratios} inserted into in \Eqs{eq:solve4m1}{eq:solve4m2} to obtain the fractions.
}
\label{fig:fracs}
\end{figure}

\begin{figure}[t]
\centering
\subfloat[]{\includegraphics[scale=.755]{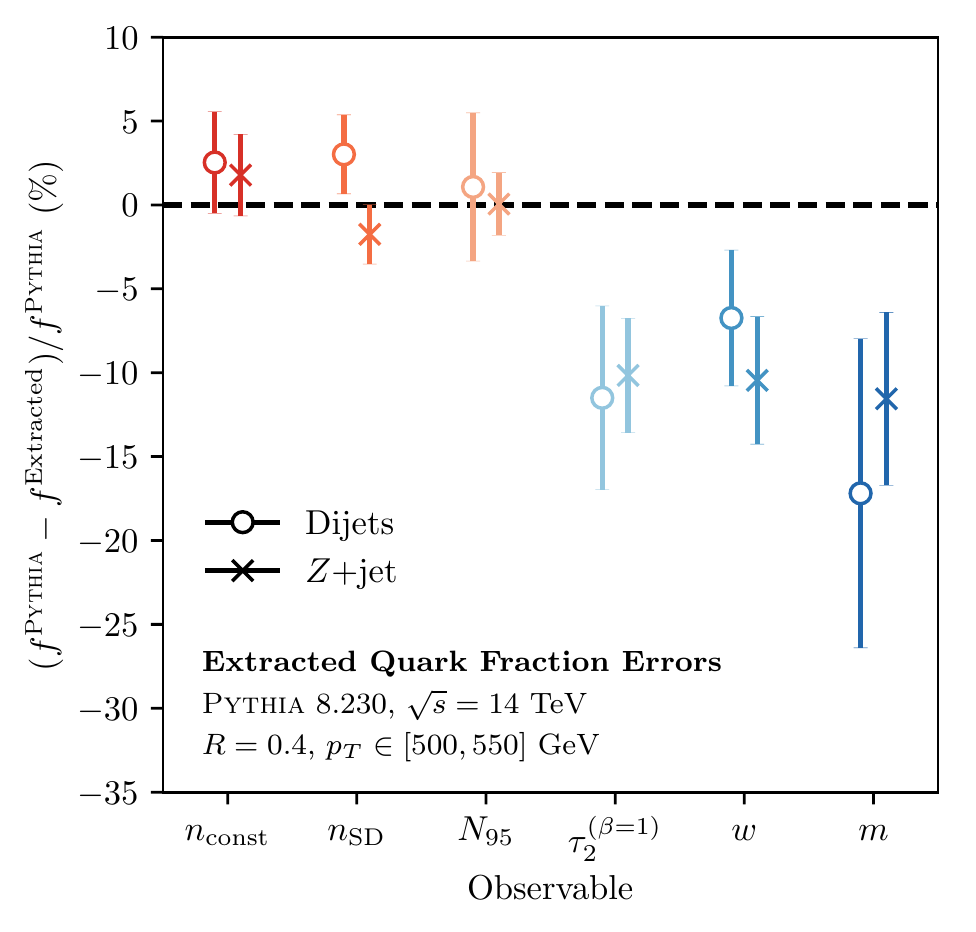}\label{fig:fracerrors-obs}}
\subfloat[]{\includegraphics[scale=.755]{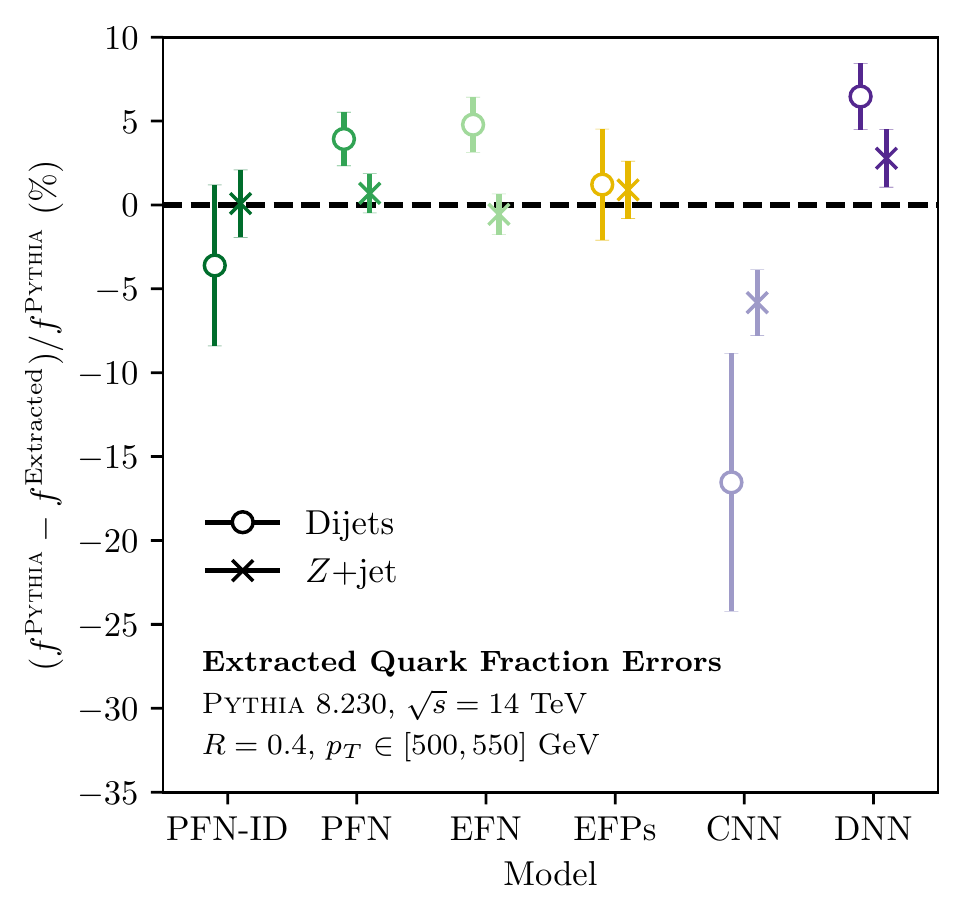}\label{fig:fracerrors-model}}
\caption{
The percent error of the extracted quark fractions (see \Fig{fig:fracs}) relative to the \pythia fractions, obtained using the (a) individual observables and (b) trained models.
By this measure, the best jet observable appears to be $N_{95}$ and the best model is the linear EFP model.
}
\label{fig:fracerrors}
\end{figure}

Using the extracted extrema of the mixed-sample log-likelihood ratio, the reducibility factors can be obtained by appropriate exponentiation.
The quark fractions can then be calculated according to \Eqs{eq:solve4m1}{eq:solve4m2}.
These are shown in \Fig{fig:fracs-obs} for the individual observables and \Fig{fig:fracs-model} for the trained models.
We see that the trained models all extract fractions largely consistent with one another and with the \pythia fractions.
The count substructure observables also extract consistent fractions, while the shape observables exhibit Casimir-scaling behavior, making them unsuitable for identifying mutually-irreducible quark and gluon jets. 
The fractions obtained from the trained models were consistently more robust to different choices of topic extraction procedures, such as the histogram binning.
Despite having little to no handle on the details of the trained models, we are able to obtain important constraints on their behavior and use them to obtain quark/gluon fractions, which are evidently insensitive to these details.

As a more quantitative measure of the quality of the extracted quark fractions, the percent error of the extracted fractions relative to the (unphysical) \pythia fractions is shown in \Figs{fig:fracerrors-obs}{fig:fracerrors-model}.
The count observables and trained models agree within several statistical uncertainties of one another and the \pythia fractions, in many cases achieving $\mathcal O(1\%)$ fidelity.
Again, we caution that the \pythia fractions solely provide a heuristic to demonstrate the performance of the method and should not be taken as fundamental to quark and gluon jets.

\subsection{Self-calibrating classifiers}
\label{sec:selfcal}

\begin{figure}[t]
\centering
\includegraphics[scale=0.85]{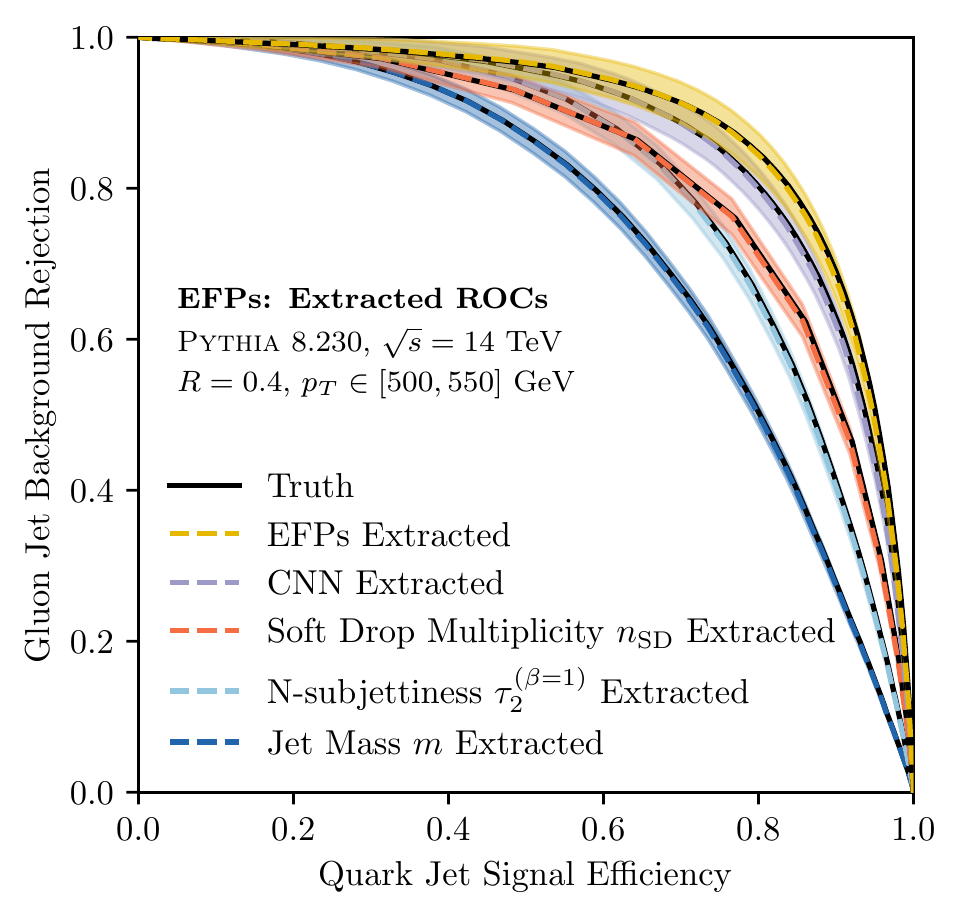}
\caption{
The ROC curves for several substructure observables and trained models using the quark fractions estimated from the EFPs.
The ``Truth'' corresponds to using the \pythia fractions to obtain the ROC curve.
We see good agreement between the data-driven ROC curves and the \pythia-labeled ROC curves.
Further, we see that the CWoLa-trained EFP classifier has effectively self-calibrated itself in this way.}
\label{fig:extractedrocs}
\end{figure}

With the quark fractions of the mixtures in hand, one immediate application is to use them to calibrate the quark/gluon classifiers, as discussed in \Sec{sec:optimal}.
Since uncalibrated classifiers can be used to obtain these fractions, this allows for self-calibrating classifiers in the CWoLa framework.
This liberates the CWoLa framework from necessarily requiring a small test set with known fractions (c.f.~\Ref{Metodiev:2017vrx}).
In the present picture, this ability to self-calibrate is conceptually clear since a sample with ``known'' fractions is equivalent to providing a definition of the underlying categories.

Beyond solely self-calibration of classifiers, the extracted fractions can be used to obtain the receiver operating characteristic (ROC) curves for other trained models or substructure observables, even those that do not themselves exhibit quark/gluon mutual irreducibility.
The extracted ROC curves of a variety of trained model and substructure observables using the EFP-extracted quark fractions are shown in \Fig{fig:extractedrocs}, with estimated uncertainty bands coming from uncertainties in the extracted fractions.
They are compared to the calibrated ROC curve using the \pythia-labeled fractions, achieving very good agreement between the two.
Note that the uncertainties are smaller for worse classifiers, which is intuitive given the limit that a perfectly-random classifier can be identified as such without any fraction information.
Overall, this concretely demonstrates that the self-calibration of CWoLa-trained classifiers can be achieved in a purely data-driven way.

\subsection{Obtaining observable distributions from extracted fractions}
\label{sec:obs}

With the reducibility factors of the mixtures, the distributions of substructure observables can be extracted for quark and gluon jets separately.
This corresponds to a direct application of the Operational Definition of quarks and gluons in \Eq{eq:opdef}.
This is similar in spirit to the procedure implemented in \Refs{Aad:2014gea,ATL-PHYS-PUB-2017-009} of using quark/gluon fractions estimated by convolving matrix elements and parton distribution functions and then solving systems of linear equations.
The key distinction is that, in our case, the fractions (and reducibility factors) themselves are data-driven.

\begin{figure}[t]
\centering
\includegraphics[scale=.66]{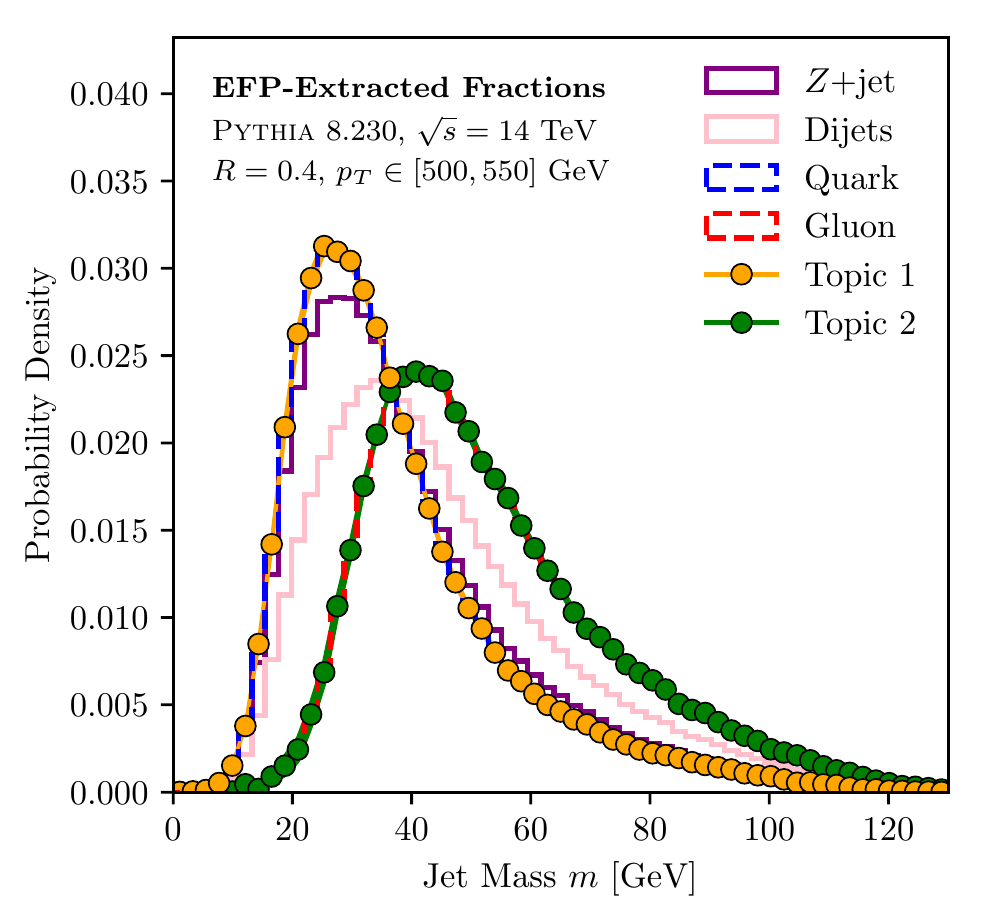}
\includegraphics[scale=.66]{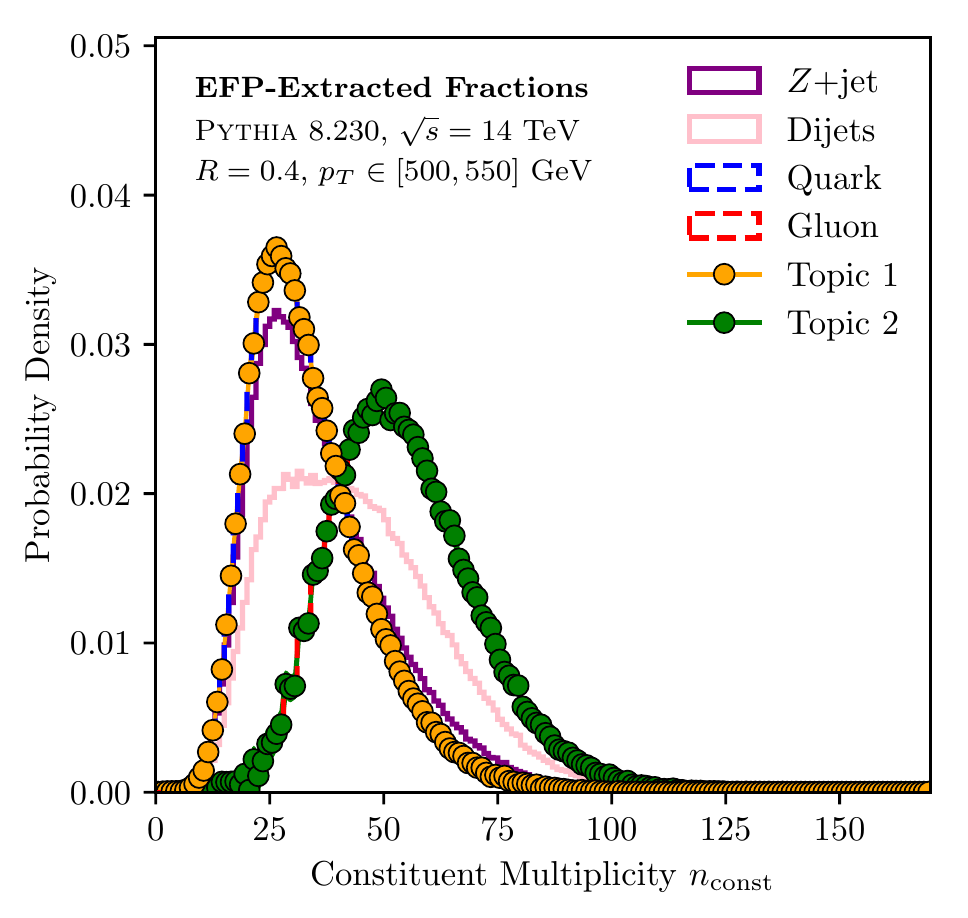}

\includegraphics[scale=.66]{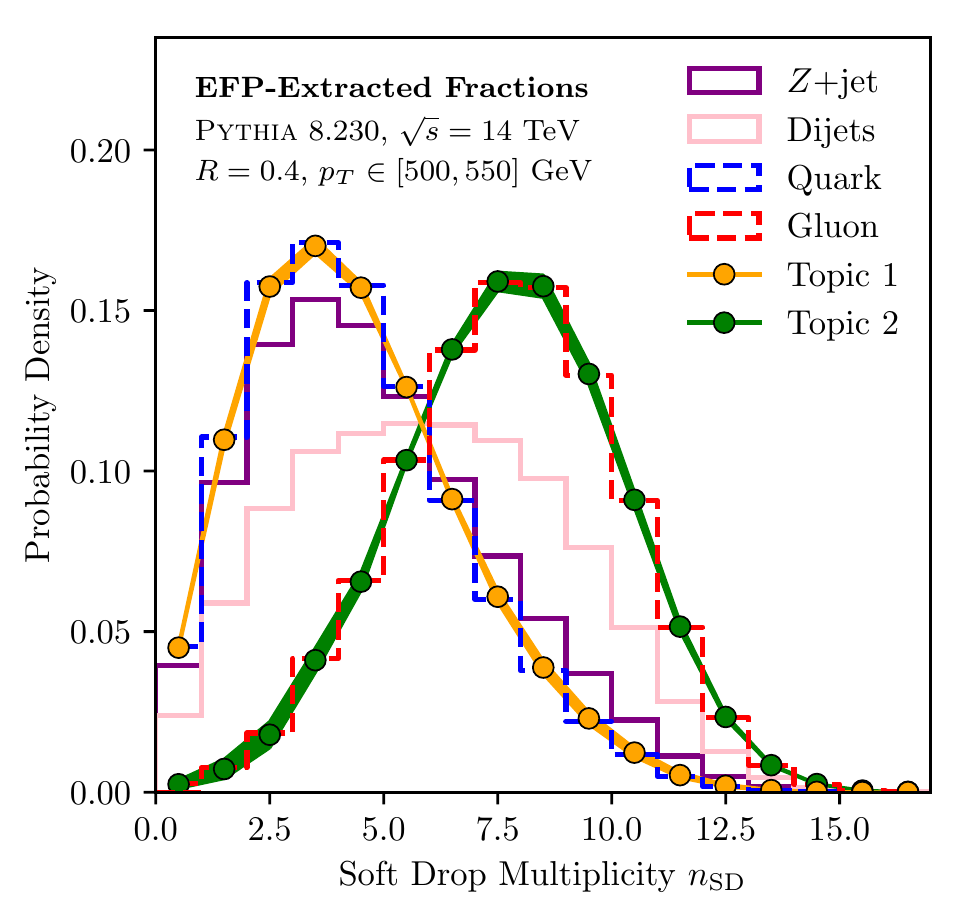}
\includegraphics[scale=.66]{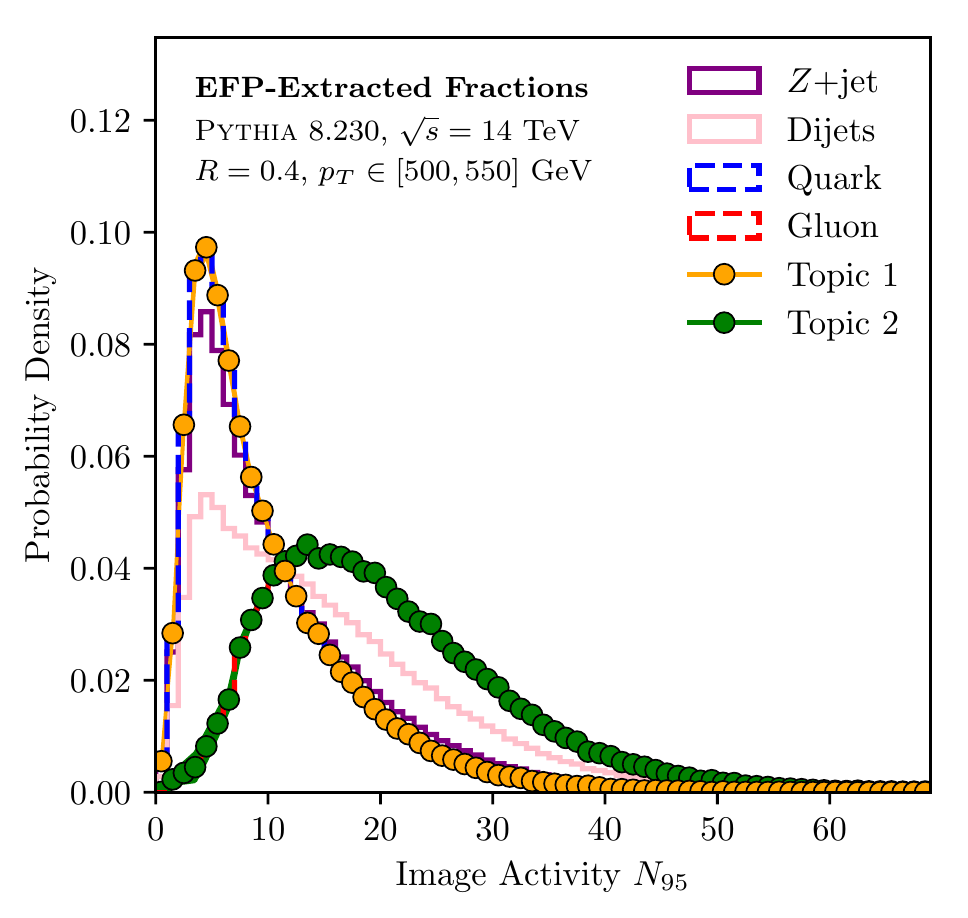}

\includegraphics[scale=.66]{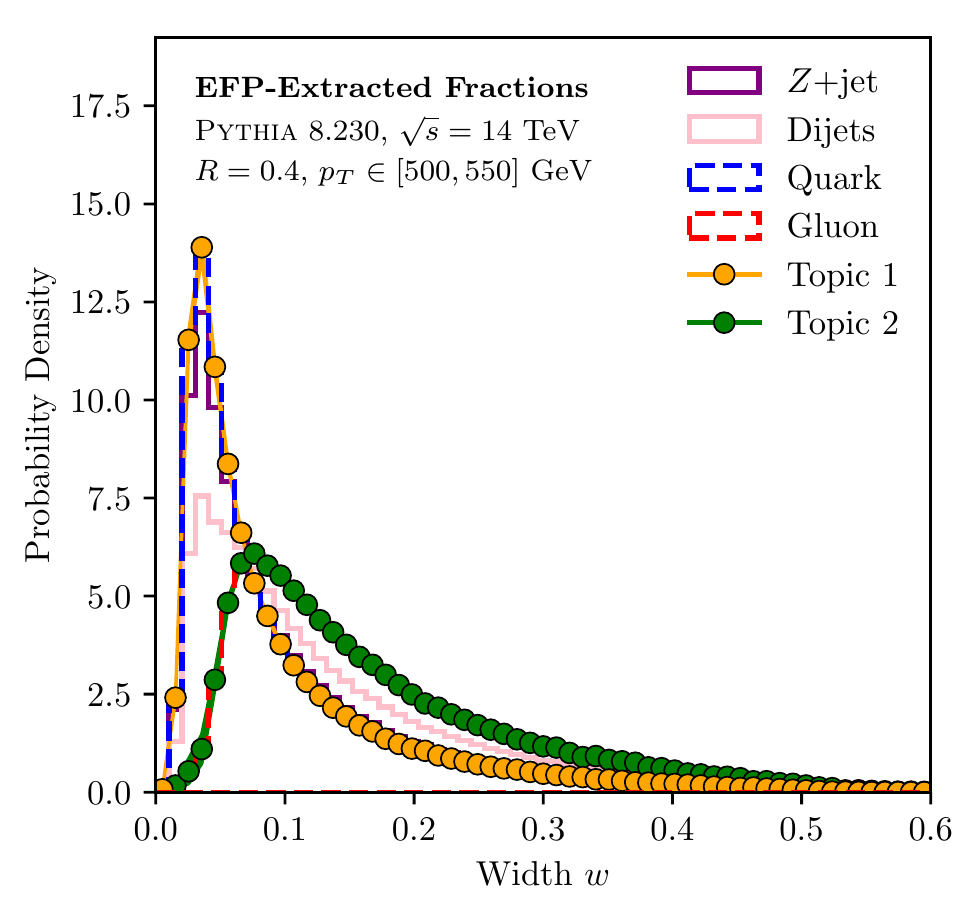}
\includegraphics[scale=.66]{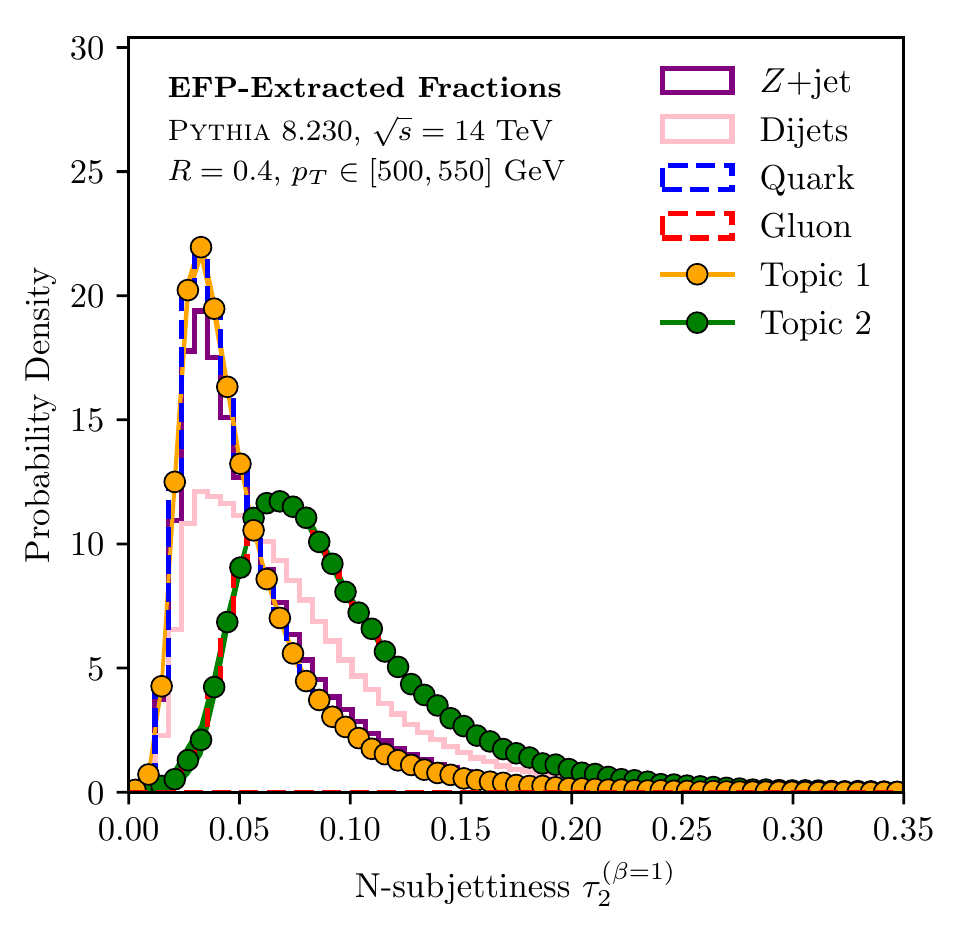}
\caption{
The distributions of the six substructure observables in the $Z+$jet sample (purple) and dijet sample (pink), with the quark and gluon distributions determined from the \pythia fractions (blue and red, respectively) and the jet topics (orange and green) using EFP-extracted reducibility factors.
We see excellent agreement between the jet topics and the \pythia-determined distributions of quark and gluon jets.
}
\label{fig:extractedhists}
\end{figure}

In \Fig{fig:extractedhists}, we use the reducibility factors defined by the EFP classifier to extract quark and gluon distributions for the six individual substructure observables. 
We see excellent agreement between the data-driven, operationally-defined quark and gluon distributions and the ones specified by the \pythia fractions.
Importantly, this procedure works for any substructure observable, even ones such as jet mass and width which do not manifest quark/gluon mutual irreducibility.

\afterpage{\clearpage}

\section{Conclusions}
\label{sec:conc}

In this paper, we provided an Operational Definition of quark and gluon jets, based solely on physical cross section measurements. 
We connected our definition to the existing CWoLa and jet topics paradigms, showing how they each fit naturally into the implementation of the definition.
Taking two mixed samples, for which there is a qualitative notion that one is more ``quark-like'' than the other, the Operational Definition returns a quantitive understanding through mutually-irreducible quark and gluons distributions.
Practically, we implemented this definition by approximating the mixed-sample likelihood ratio, relating it to the pure quark/gluon likelihood ratio, and finding its extrema to determine mixed-sample reducibility factors.
With the reducibility factors in hand, the quark fractions for the mixed samples can be readily obtained.
In a broad sense, our Operational Definition harmonizes with the statistical picture of jet samples at colliders, where individual jets do not carry intrinsic flavor labels and one only ever has access to mixed samples in data.

To illustrate the power of the Operational Definition, we tested it in the realistic context of $Z$+jet and dijet processes.
We applied our quark/gluon jet definition to twelve different observables:  six individual substructure observables, and six trained machine learning models which distilled a huge feature space down to a single optimal observable.
The six individual observables naturally fall into two categories, count and shape observables, and we confirmed that the count observables yield much more accurate quark fractions (relative to a \pythia baseline).
With the minor exception of the CNN, the machine learning models all did well at extracting the fractions.
While the performance of the best individual observable ($N_{95}$) and the best machine learning model (linear EFPs) were comparable, the machine learning models were overall more robust to changes in histogram binning and to the technique used for determining the reducibility factors; this in turn contributes to the robustness of the Operational Definition.
Having determined the quark fractions, we extracted pure quark and gluon distributions for various jet substructure observables.
Crucially, this worked even for observables that do not exhibit quark/gluon mutually irreducibility, as long as the observable used to extract the fractions does.
Additionally, we demonstrated that CWoLa classifiers could be self calibrated using fractions obtained from an uncalibrated classifier, thereby removing a potential hurdle in using CWoLa in practice.

The techniques in this paper represent a novel use of classification in particle physics.
Instead of tagging quark and gluon jets, we used a CWoLa-trained deep learning classifier to approximate the full mixed-sample likelihood ratio.
This is in the same spirit of recent work on deep learning~\cite{Andreassen:2018apy,Komiske:2017ubm,Komiske:2017aww,Chang:2017kvc,Roxlo:2018adx,deOliveira:2017pjk,Paganini:2017hrr,Paganini:2017dwg,DAgnolo:2018cun}, where the ``black box'' nature of the trained model is not of central importance to the success or understanding of the method.
No longer is the output of a neural network viewed as an arbitrary quantity used only for discrimination, but rather as a robust approximation to the likelihood ratio, which turns out also to be optimal for extracting categories from the data.
Surprisingly, while individual quark and gluon jets cannot be tagged perfectly, we were able to use a data-driven classifier to extract the full quark and gluon distributions of an observable to percent-level accuracy.
This approach paves the way for fully data-driven collider physics, making use of machine learning techniques trained directly on data while producing results insensitive to the details of the ``black box''.

We conclude by discussing potential extensions of the methods used in this paper.
As mentioned in \Sec{sec:topicwola}, a key concern in jet tagging is sample dependence, i.e.\ whether a ``quark jet'' in one sample is the same as a ``quark jet'' in another.
While the Operational Definition sidesteps the issue of sample dependence in the case of two mixed samples, it is natural to ask what happens with three or more mixed samples.
Concretely, once the Operational Definition is applied to two mixed jet samples, one can ask the degree to which a third sample $M$ is explained by the existing quark and gluon distributions.
It turns out that there is a unique and well-defined generalization of the reducibility factor, discussed in \Ref{katz2017decontamination}, that precisely captures this notion and yields a quantifiable notion of sample dependence:
\begin{equation}\label{eq:other}
\kappa \equiv \max_{f_q,\,f_g}\{f_q + f_g \,|\,\exists \text{ dist. } p_o(\O) \text{ s.t. } p_{M}(\O) = f_q p_q(\O) + f_g p_g(\O) + (1-f_q-f_g)p_o(\O)\},
\end{equation}
where $0\le f_q,f_g\le 1$ and $f_q+f_g\le1$.
In \Eq{eq:other}, $\kappa$ is the maximum amount of $M$ explainable by the quark and gluon distributions, requiring minimal addition of an ``other'' distribution $p_o(\mathcal O)$.
Understanding sample dependence is a general challenge, even with parton-shower-extracted templates, so it is gratifying that our framework naturally suggests a tool to address this problem.
Sample dependence can also be studied by directly comparing the quark and gluon jet definitions provided by different pairs of jet samples ($Z$+jet, dijets, $\gamma$+jet, etc.)\ at different transverse momenta and jet radii.
We leave explorations of these important ideas, as well as more detailed optimizations of the method, to future work.

Extending this thinking, one might attempt to provide a concrete jet flavor definition beyond the two-category case of quarks and gluons.
For instance, while the difference in radiation patterns between different-flavor light-quark jets is much smaller than between quark and gluon jets, it may be possible to use the techniques described in this paper to define differently-flavored quark jets.
The subtle difference in radiation patterns between different light-quark has been studied in the context of jet charge observables in \Ref{Krohn:2012fg} and in the context of machine learning in \Ref{Fraser:2018ieu}.
To use our methods in this case would require advances in multiple-category CWoLa and jet topics, though the conceptual underpinnings would be the same as for the two-category case studied here.
Further, one could extend such a definition to provide well-defined jet flavor definitions for a variety of other boosted hadronic objects, potentially including subtle distinctions like longitudinal versus transverse polarization of boosted $W/Z$ bosons.
More broadly, the concept of mutual irreducibility as a means of defining categories may find additional applications in high-energy physics due to its utility in disentangling overlapping distributions using pure phase space signatures.

\acknowledgments 
The authors would like to thank Jonathan Butterworth, Philip Harris, Andrew Larkoski, Benjamin Nachman, Matthew Schwartz, and Clayton Scott for insightful comments and helpful discussions.
This work was supported by the Office of Nuclear Physics of the U.S. Department of Energy (DOE) under grant DE-SC-0011090 and the DOE Office of High Energy Physics under grant DE-SC-0012567.
Cloud computing resources were provided by a Microsoft Azure for Research Award.

\appendix

\section{Theoretical exploration of Casimir- and Poisson-scaling observables}
\label{sec:explore}

In this appendix, we explore the Operational Definition of quark and gluon jets in the leading-logarithmic (LL) limit, focusing on two theoretically-tractable classes of jet observables: Casimir-scaling and Poisson-scaling observables.
Though we only work to lowest non-trivial order, these calculations demonstrate that our framework for defining quark and gluon jets is suitable to theoretical exploration in addition to practical experimental implementation.
In the LL limit of perturbative QCD, quarks and gluons differ in their emission profiles only by their color charges: $C_F=4/3$ for quarks and $C_A=3$ for gluons.
Thus, in the LL limit, quarks and gluons are well-defined (at least at the parton level), providing a simplified context to explore the Operational Definition.
We find different non-zero quark/gluon reducibility factors for Casimir-scaling and Poisson-scaling observables, substantiating the need to use a richer space of jet substructure observables to approximate the full likelihood ratio.

Casimir-scaling observables include common jet substructure observables, such as the jet mass $m$ or IRC-safe angularities~\cite{Berger:2003iw,Almeida:2008yp,Ellis:2010rwa,Larkoski:2014uqa,Larkoski:2014pca}, that are dominated at LL accuracy by a single hard emission.
Their cumulative distributions satisfy $\Sigma_g (m)= \Sigma_q(m)^{C_A/C_F}$, where $p_i(m) = d\Sigma_i/dm$.
Solely using this scaling property, the quark/gluon reducibility factors of Casimir-scaling observables are:
\begin{align}\label{eq:kqgcas}
&\kappa_{qg}^\text{Cas.} = \min_m\frac{p_q(m)}{p_g(m)} = \min_m\frac{\frac{d\Sigma_q}{dm}}{\frac{C_A}{C_F} \Sigma_q^{C_A/C_F - 1}\frac{d\Sigma_q}{dm}}  = \frac{C_F}{C_A} \min_m\Sigma_q^{1-C_A/C_F} = \frac{C_F}{C_A}, \\
&\kappa_{gq}^\text{Cas.} = \min_m\frac{p_g(m)}{p_q(m)} = \min_m\frac{\frac{C_A}{C_F} \Sigma_q^{C_A/C_F - 1}\frac{d\Sigma_q}{dm}}{\frac{d\Sigma_q}{dm}}  = \frac{C_A}{C_F} \min_m\Sigma_q^{C_A/C_F-1} = 0, \label{eq:kgqcas}
\end{align}
where $C_A/C_F > 1$ and $\min_m\Sigma_i(m)=0$ have been used to obtain the last equality.
These results are universal to all Casimir-scaling observables and are independent of the remaining details of the observables at LL accuracy.

\begin{figure}[t]
\centering
\subfloat[]{\includegraphics[scale=0.75]{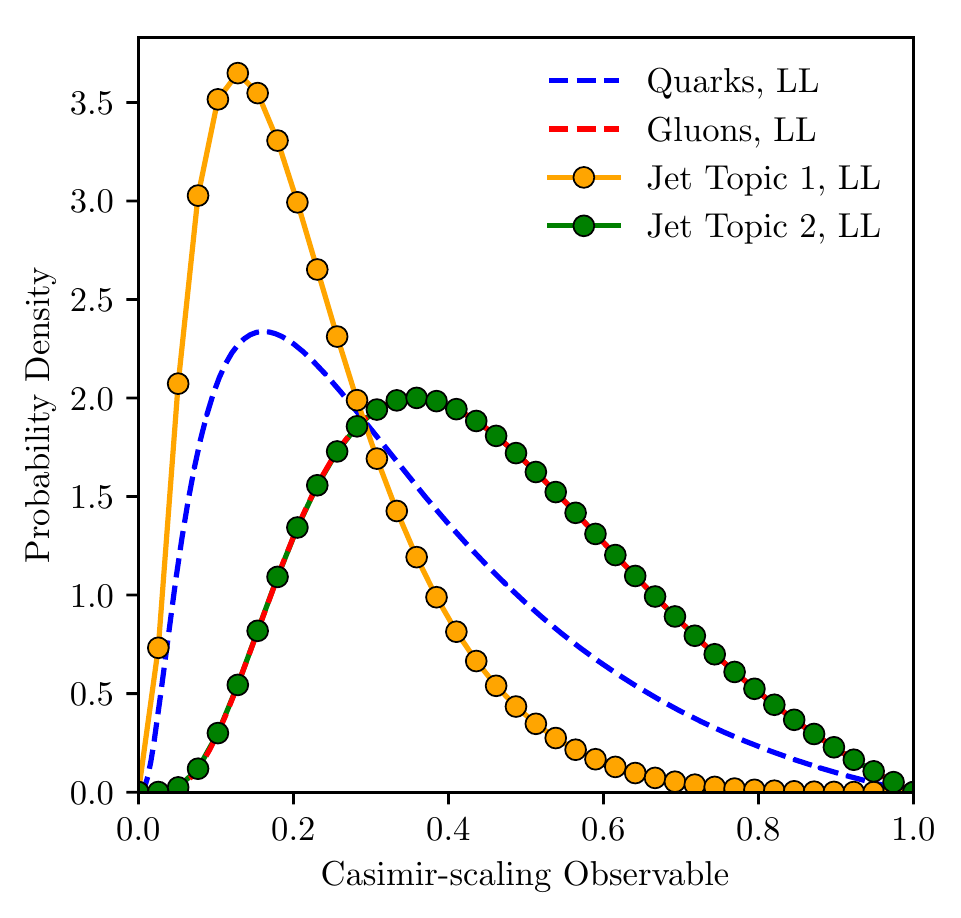}\label{fig:CStopics}}
\subfloat[]{\includegraphics[scale=0.75]{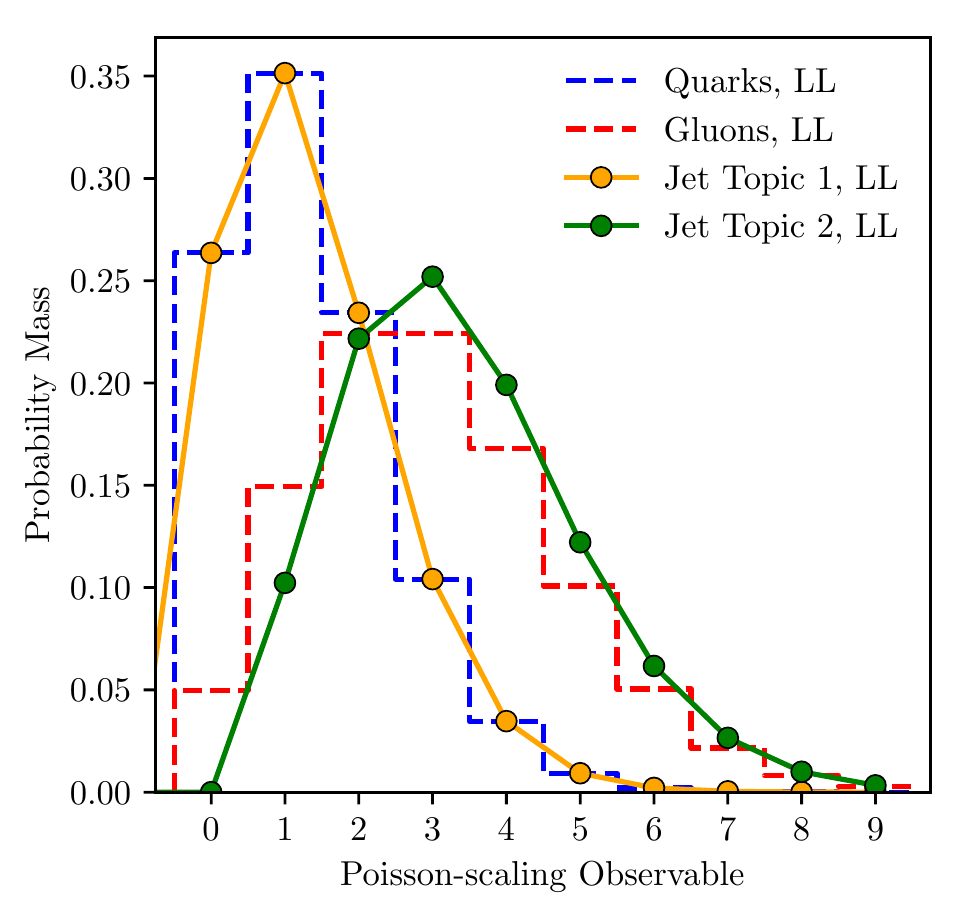}\label{fig:PStopics}}
\caption{
Quark and gluon distributions at LL accuracy for (a) Casimir-scaling and (b) Poisson-scaling observables, together with the corresponding jet topics.
The reducibility of the quark Casimir-scaling distribution and the gluon Poisson-scaling distribution are evident.
While neither of these observables individually results in mutually irreducible quarks and gluons, considering them jointly does.
}
\label{fig:LLtopics}
\end{figure}

The non-zero reducibility factor in \Eq{eq:kqgcas} indicates that quark and gluon jets are not mutually irreducible in the space of Casimir-scaling observables.
In particular, the quark distribution of any Casimir-scaling observable is a mixture of the (irreducible) gluon distribution and some other distribution, as shown in \Fig{fig:CStopics}.
Note that this {\it does not} imply that quark jets are fundamentally reducible, since this is just a property derived from Casimir-scaling observables in the LL limit.
That said, as noted at the end of \Sec{sec:opdef}, if \Eq{eq:kqgcas} were fundamental to quark and gluon jets, one could simply include this reducibility factor in the Operational Definition.

We next consider Poisson-scaling observables, which count the number of perturbative emissions and have qualitatively different quark-gluon reducibility factors.
One example is the soft drop multiplicity $n_{\rm SD}$~\cite{Frye:2017yrw}, which counts the number of emissions restricted to a certain phase space region.
At LL, Poisson-scaling observables are distributed according to Poissonian distributions with means $C_F \lambda$ for quarks and $C_A \lambda$ for gluons, where $\lambda$ is a constant proportional to the area of the emission plane that is counted.
The quark-gluon reducibility factors corresponding to these distributions are then:
\begin{align}\label{eq:kqgpois}
&\kappa_{qg}^\text{Pois.} = \min_n\frac{p_q(n)}{p_g(n)} = \min_n\frac{(C_F \lambda)^n e^{- C_F \lambda}}{(C_A \lambda)^n e^{- C_A \lambda}} =e^{-(C_F - C_A)\lambda}  \min_n \left(\frac{C_F}{C_A}\right)^n = 0,\\
&\kappa_{gq}^\text{Pois.} = \min_n\frac{p_g(n)}{p_q(n)} = \min_n\frac{(C_A \lambda)^n e^{- C_A \lambda}}{(C_F \lambda)^n e^{- C_F \lambda}} = e^{-(C_A-C_F) \lambda} \min_n \left(\frac{C_A}{C_F}\right)^n = e^{-(C_A-C_F)\lambda},\label{eq:kgqpois}\end{align}
since $C_A/C_F > 1$ and $n$ can take any non-negative integer value.

Evidently, Poisson-scaling observables display the {\it opposite} behavior of Casimir-scaling observables: the gluon distribution is a mixture of the (irreducible) quark distribution and some other distribution, as shown in \Fig{fig:PStopics}.
Further, the reducibility factor is not universal to all Poisson-scaling observables but rather depends exponentially on the parameter $\lambda$.
Though $\lambda\sim\mathcal O(1)$ was considered in \Ref{Frye:2017yrw}, perturbative QCD allows for arbitrarily large $\lambda$ by counting emissions in larger and larger regions.
As $\lambda$ increases, the reducibility factor falls to zero much more quickly than the overlap in the distributions decreases, and thus quark and gluon jets rapidly approach mutual irreducibility.
While perturbative control is lost for large $\lambda$ due to non-perturbative effects, considering this limit suggests that there is no fundamental impediment to the mutual irreducibility of quarks and gluons from the perspective of perturbative QCD, at least at LL accuracy.

From these two classes of observables, we see that enriching the feature space beyond individual Casimir-scaling and Poisson observables to $\mathcal O = \{m,n_{\rm SD}\}$ yields $\kappa_{qg} = \kappa_{gq} = 0$ for the combined feature space in the LL limit.
This benefit of using a rich feature space motivates our approach of training data-driven classifiers on complete substructure information to probe the full quark/gluon jet likelihood ratio, rather than relying on individual specially-crafted substructure observables.

\section{Details of observables and machine learning models}
\label{sec:train}

In this appendix, we give details for the jet substructure study in \Sec{sec:qgex}, describing the observables, machine learning models, and model training used.

For the individual substructure observables, three of them use custom implementations: constituent multiplicity $n_\text{const}$, image activity $N_{95}$~\cite{Pumplin:1991kc} (number of pixels in a $33\times33$ jet image containing 95\% of the $p_T$), and jet mass $m$.
The remaining three observables are computed using \textsc{FastJet contrib} 1.033~\cite{fjcontrib}.
The \textsc{RecursiveTools} 2.0.0-beta1 module is used to calculate soft drop multiplicity $n_{\rm SD}$~\cite{Frye:2017yrw} with parameters $\beta=-1$, $z_\text{cut}=0.005$, and $\theta_\text{cut}=0$.
The \textsc{Nsubjettiness} 2.2.4 module is used to calculate the $N$-subjettiness~\cite{Thaler:2010tr,Thaler:2011gf} observables $\tau_N^{(\beta)}$ with $k_T$ axes as recommended in \Ref{Datta:2017rhs}, in particular $\tau_2^{(\beta=1)}$ and jet width $w$ (implemented as $\tau_1^{(\beta=1)}$).

For our trained models, we use several different jet representations and machine learning architectures.
In reverse order compared to \Tab{tab:obsandmodels}, they are:
\begin{itemize}

\item \textbf{DNN}: The $N$-subjettiness basis~\cite{Datta:2017rhs} is a phase space basis in the sense that $3K-4$ independent $N$-subjettiness observables map non-linearly onto $K$-body phase space.
We use 20-body phase space consisting of the following set of $N$-subjettiness basis elements:
\begin{equation}
\left\{\tau_1^{(1/2)},\,\tau_1^{(1)},\,\tau_1^{(2)},\tau_2^{(1/2)},\,\tau_2^{(1)},\,\tau_2^{(2)},\,\ldots,\,\tau_{K-2}^{(1/2)},\,\tau_{K-2}^{(1)},\,\tau_{K-2}^{(2)},\tau_{K-1}^{(1/2)},\,\tau_{K-1}^{(1)}\right\},
\end{equation}
i.e.\ $\tau_N^{(\beta)}$ with $N\in\{1,\ldots,19\}$ and $\beta\in\{1/2,1,2\}$, except $\tau_{19}^{(2)}$ is absent, all computed using the \textsc{Nsubjettiness} 2.2.4 module of \textsc{FastJet contrib} 1.033.
A DNN consisting of three 100-unit fully-connected layers and a 2-unit softmaxed output was trained on the $N$-subjettiness basis inputs.

\item \textbf{CNN}: The jet images approach~\cite{Cogan:2014oua} treats calorimeter deposits as pixel intensities and represents the jet as an image. 
Convolutional neural networks (CNNs) are the typical model of choice when learning from such a representation, and have been successfully implemented for quark/gluon discrimination~\cite{Komiske:2016rsd}, $W$ tagging~\cite{deOliveira:2015xxd}, and top tagging~\cite{Baldi:2016fql,Guest:2016iqz}.
We calculate $33\times33$ jet images spanning $2R\times2R$ in the rapidity-azimuth plane.
In the language of \Ref{Komiske:2016rsd}, we formulate ``color'' jet images with two channels: the $p_T$ per pixel and the multiplicity per pixel.
Images were standardized by subtracting the mean and dividing by the per-pixel standard deviation of the training set.

A CNN architecture similar to that used in \Ref{Komiske:2016rsd} was employed: three convolutional layers with 48, 32, and 32 filters and filter sizes of $8\times 8$, $4\times 4$, and $4\times 4$, respectively, followed by a 128-unit dense layer. 
Maxpooling of size $2\times2$ was performed after each convolutional layer with a stride length of 2.
The dropout rate was taken to be 0.1 for all convolutional layers and was not used for the dense layer.

\item \textbf{EFPs}: The Energy Flow basis~\cite{Komiske:2017aww} is a linear basis for IRC-safe observables in the sense that any IRC-safe observable is arbitrarily well approximated by a linear combination of Energy Flow Polynomials (EFPs).
As a result of this remarkable property, linear methods can be used for classification and regression and are highly competitive with modern machine learning methods.
The \href{https://energyflow.network}{\texttt{EnergyFlow}} 0.8.2 package~\cite{energyflow} was used to compute EFPs up to $d\le7,\,\chi\le3$ with $\beta=0.5$ using the normalized default hadronic measure.
This yields 996 EFPs in total, including the trivial constant EFP.
This set was used to train a Fisher's Linear Discriminant model with scikit-learn~\cite{scikit-learn}.

\item \textbf{EFN, PFN, PFN-ID}:
Various particle-level network architectures have been proposed to take advantage of the structure of events or jets as sequences of vectors~\cite{Louppe:2017ipp,Andreassen:2018apy,Butter:2017cot,Cheng:2017rdo,Egan:2017ojy,Komiske:2018cqr}.
We choose to focus on the Energy Flow Networks (EFNs) recently introduced in \Ref{Komiske:2018cqr} and shown to be competitive with other particle-level models.
The EFN architecture is designed to have the properties desirable of a model that takes jet constituents as inputs: it is able to handle variable length lists but, critically, is manifestly symmetric under permutations of the elements in the input.
The inputs to an EFN are lists of particles, where a particle is described by its energy fraction, rapidity, and azimuthal angle (the latter two translated to the origin according to the $E$-scheme jet axis).
EFNs construct an internal latent representation of the jet using the particle-level inputs, weighting each particle's contribution by its energy fraction in order to ensure the IRC safety of the internal observables, and then combine the internal jet observables using a DNN backend.
The \href{https://energyflow.network}{\texttt{EnergyFlow}} package contains an implementation of EFNs.

The EFN architecture can be generalized to learn potentially IRC-unsafe internal observables.
This variant is termed a Particle Flow Network (PFN), which can easily incorporate additional particle features such as flavor information; see \Ref{Komiske:2018cqr} for a more thorough discussion. 
In addition to the IRC-safe EFN, our study uses a PFN with only kinematic inputs, and a PFN-ID with both kinematic and particle flavor (or ID) information.
For each network, the per-particle frontend subnetwork has three fully-connected 100-unit layers corresponding to an internal latent representation of 100 jet observables, and the per-jet backend has three fully-connected 100-unit layers that combines the internal latent observables.
The EFN, PFN, and PFN-ID networks differ only in their inputs and whether the energy fractions are used as weights for the internal sum over particles (for the EFN) or passed to the frontend subnetwork (for the PFN and PFN-ID).

\end{itemize}

All of the above models (excepting the linear EFPs) were implemented and trained using Keras~\cite{keras} with the TensorFlow~\cite{tensorflow} backend.
Training/validation and test datasets were each constructed using 500,000 events for each jet sample being considered.
The training/validation dataset is further divided with 90\% used for training and the remaining 10\% used for validation. 
Properties common to all networks were the use of ReLU activations~\cite{nair2010rectified} for each non-output layer, a 2-unit softmaxed output layer, He-uniform initialization~\cite{heuniform} of the model weights, the categorical crossentropy loss function, the Adam optimization algorithm~\cite{adam}, a learning rate of 0.001, and a patience parameter of 10 epochs monitoring the validation loss.
Models are trained 25 times, making use of different random weight initializations, and the best one is selected according to the maximum Area Under the (mixed sample ROC) Curve.
The hyperparameters of each model were not optimized for either classification performance or accuracy of the ultimately extracted fractions but rather are demonstrative of typical performance that can be achieved.
Practical users of the Operational Definition should tune the hyperparameters for their own purpose.

Finally, it should be noted that other data-driven criteria can be used to select optimal trained models, though we do not explore this further here.
One idea is that since the regions of the ROC curve that are relevant for topic extraction are those with very low and very high signal efficiency, in practice it may be beneficial to optimize training for these regions directly.
A method for optimizing loss-function based training by operating point is described in \Ref{rocopt}, and it would be fascinating to explore this for training better models for topic extraction.

\section{Sample dependence in parton shower events}
\label{sec:sampledependence}

\begin{figure}[p]
\centering
\includegraphics[scale=.66]{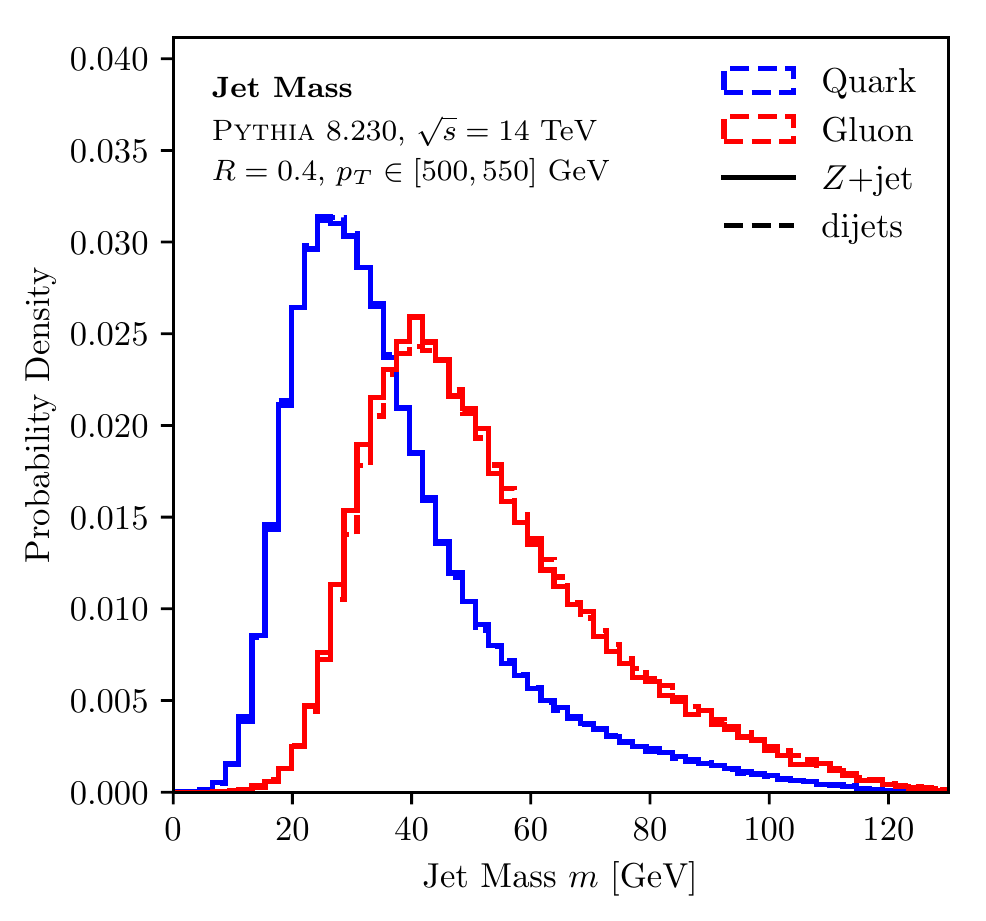}
\includegraphics[scale=.66]{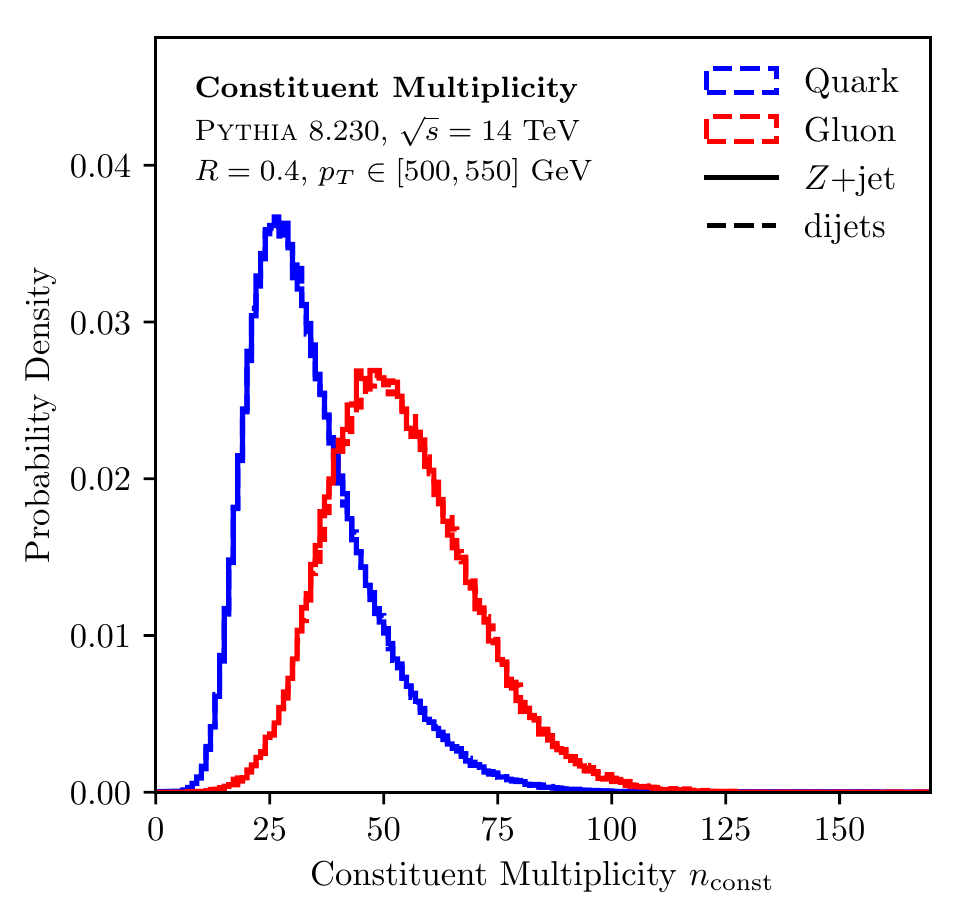}

\includegraphics[scale=.66]{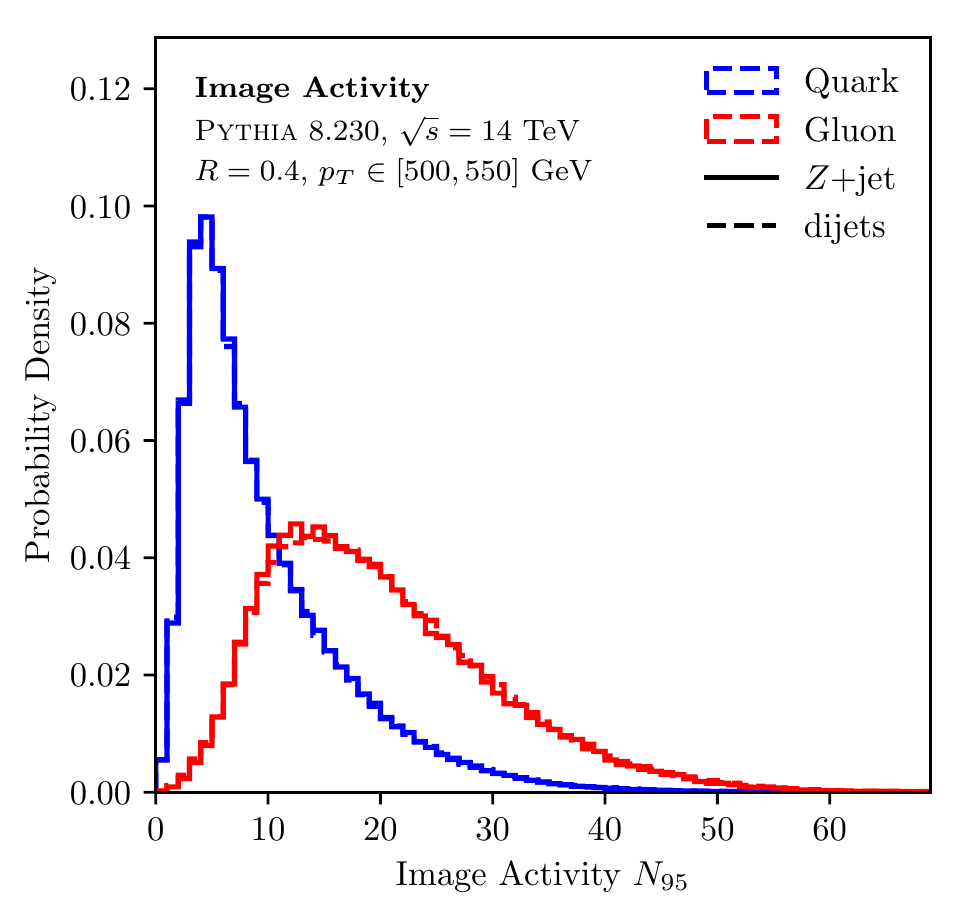}
\includegraphics[scale=.66]{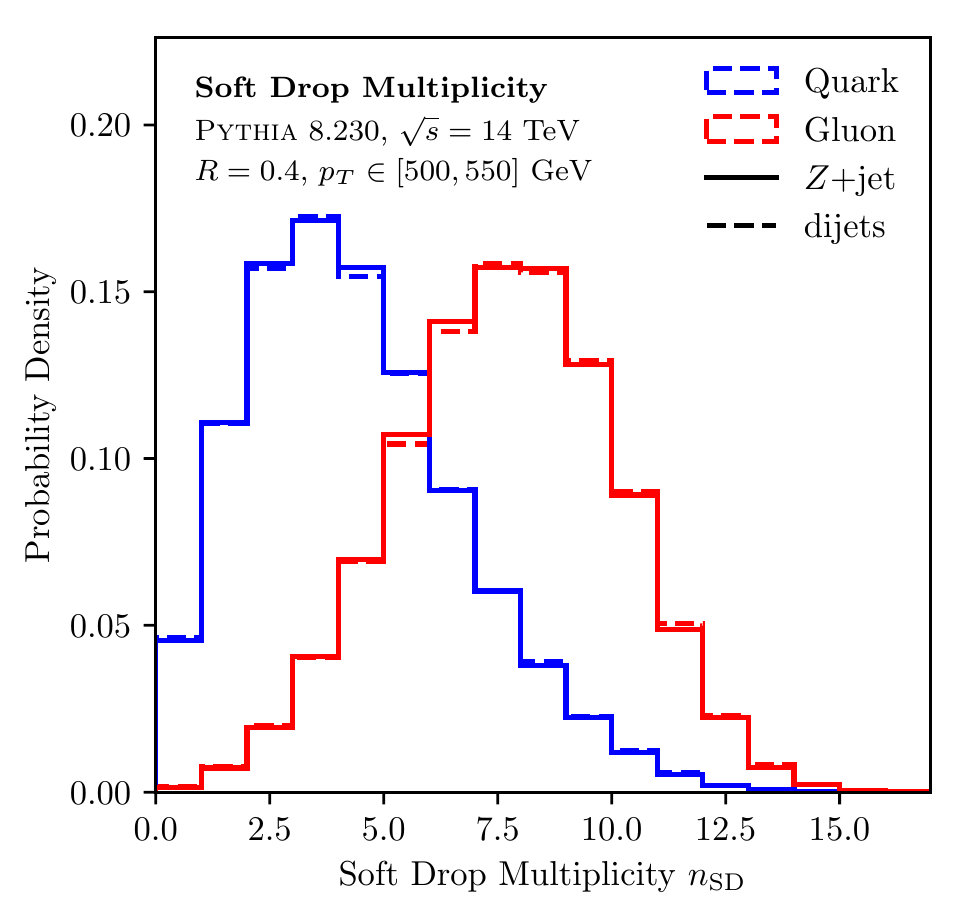}

\includegraphics[scale=.66]{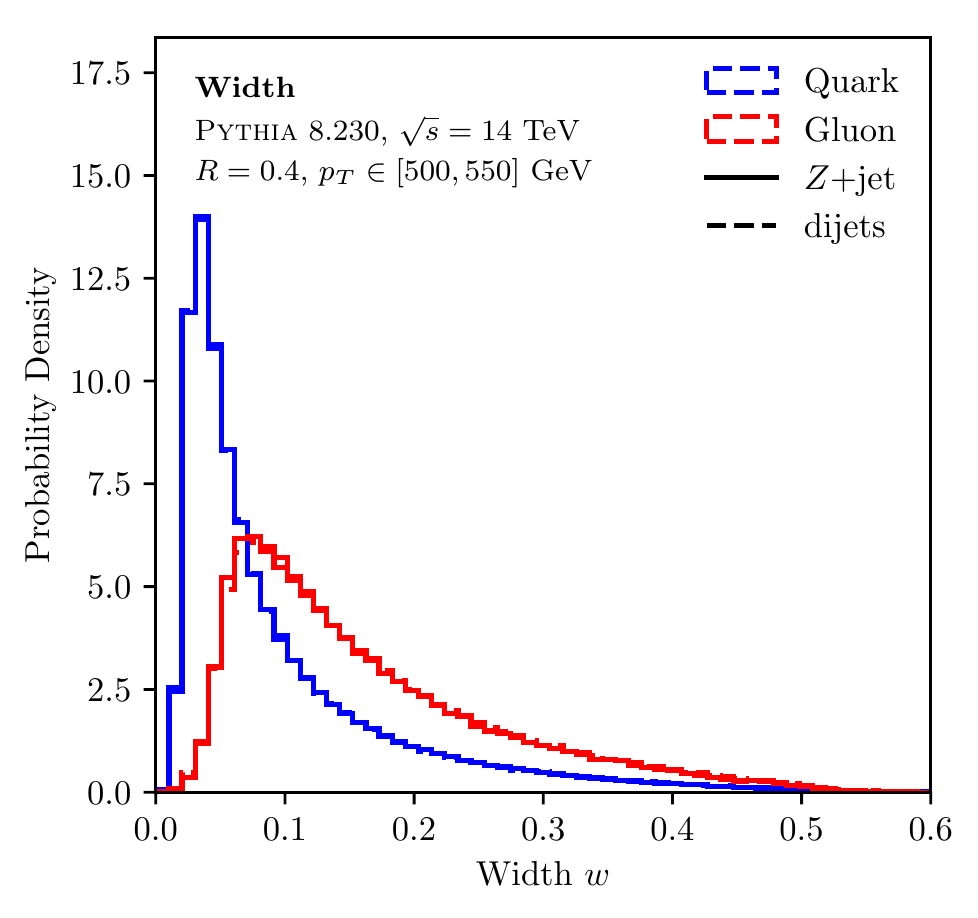}
\includegraphics[scale=.66]{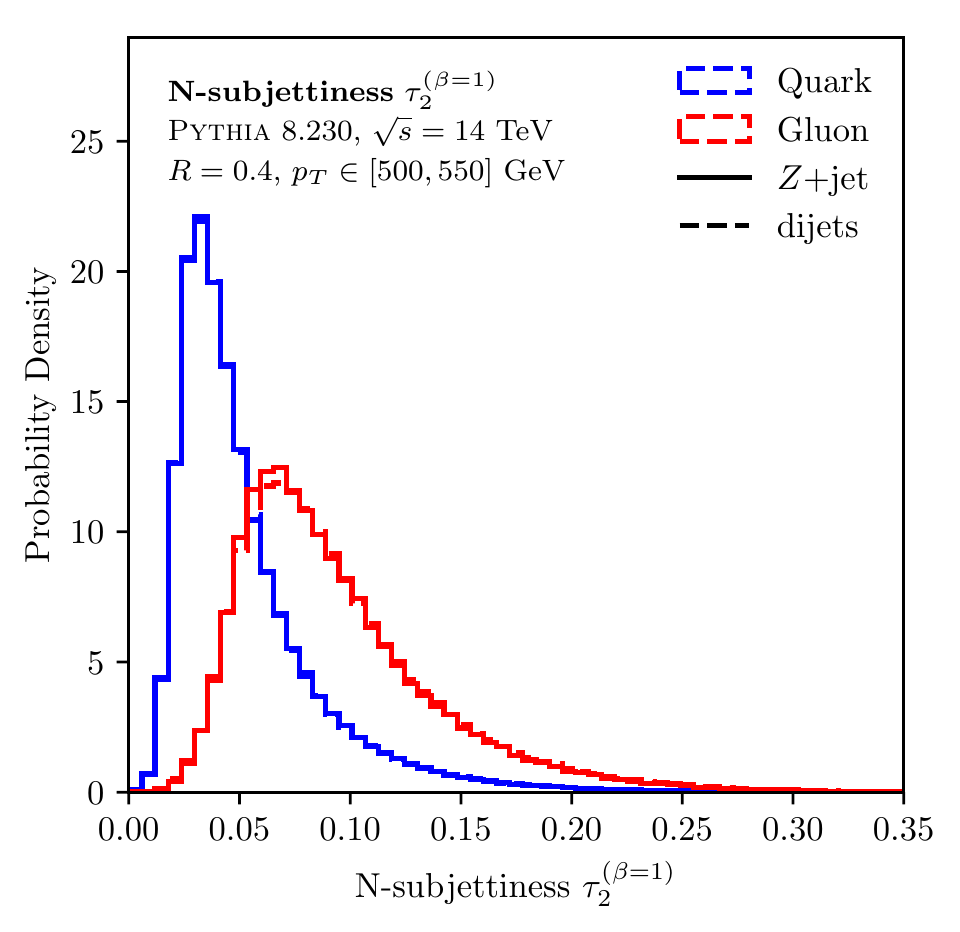}

\caption{Distributions for the six individual jet observables for $Z$+jet quarks (solid blue), $Z$+jet gluons (solid red), dijet quarks (dashed blue), and dijet gluons (dashed red). 
That the quark and gluon histograms for the two different samples are remarkably similar for this array of observables indicates a high degree of sample independence, at least for the notion of quarks and gluons in \pythia.}
\label{fig:qgobshists}
\end{figure}

In this appendix, we do a basic study of sample dependence of \pythia-labeled quark and gluon jets arising from the $Z+$jet and dijets processes.
While this is largely tangential to the main direction of the paper, it lends evidence that our case study is not far from the limit of factorized and universal notions of ``quark'' and ``gluon'' jets.
Of course, these conclusions are limited by the fact that they come from jets generated in \pythia, which itself relies on notions of factorization in its generation process.
A study of these effects in data would be an important addition to our understanding of sample independence and factorization more broadly.
We leave a study using our flavor definition to probe sample dependence in a more realistic collider setting to future work.

\begin{figure}[p]
\centering
\includegraphics[scale=.66]{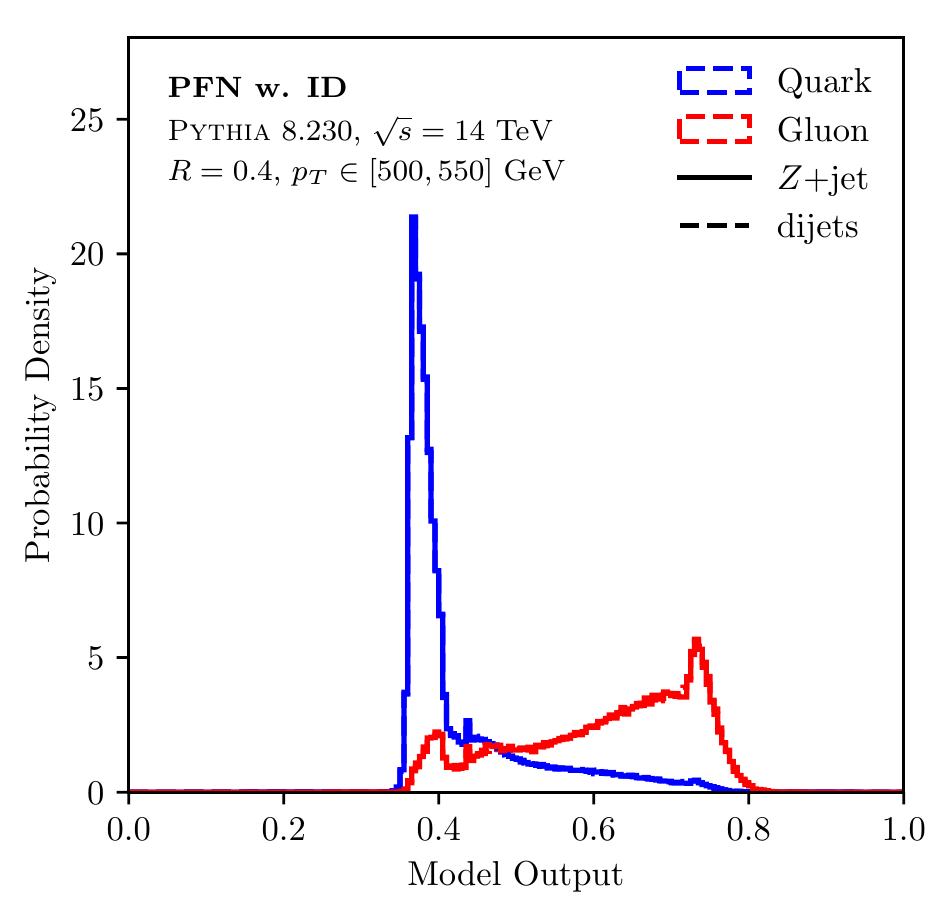}
\includegraphics[scale=.66]{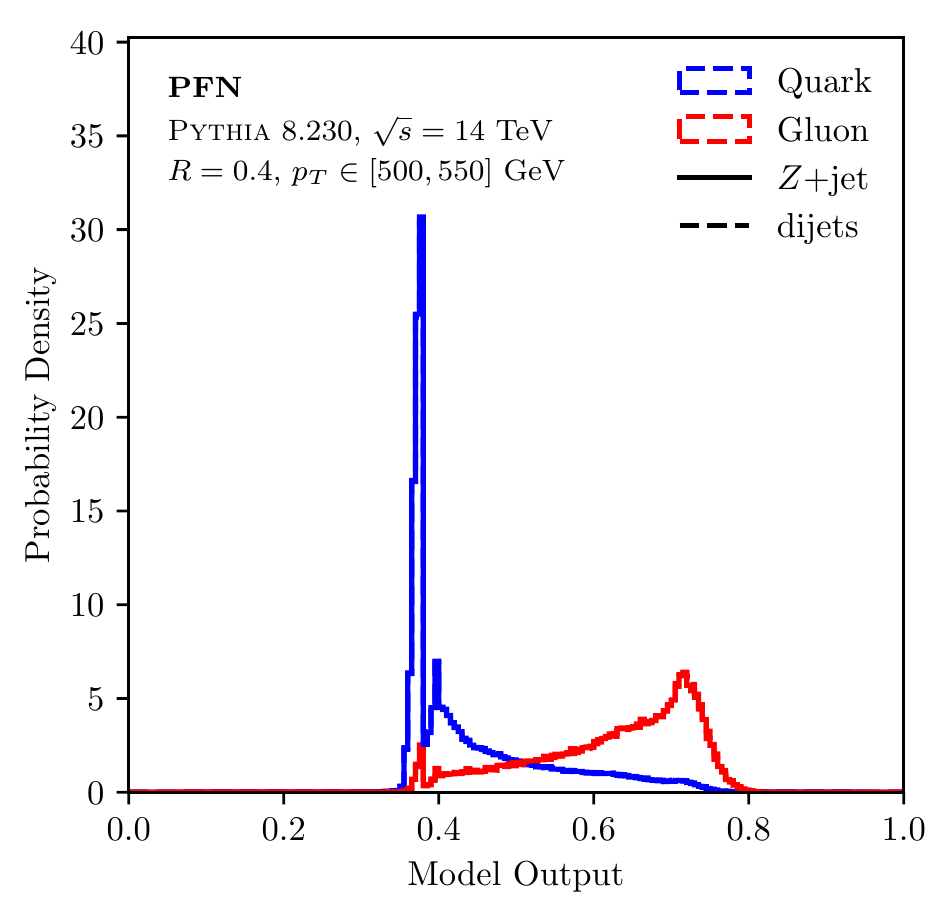}

\includegraphics[scale=.66]{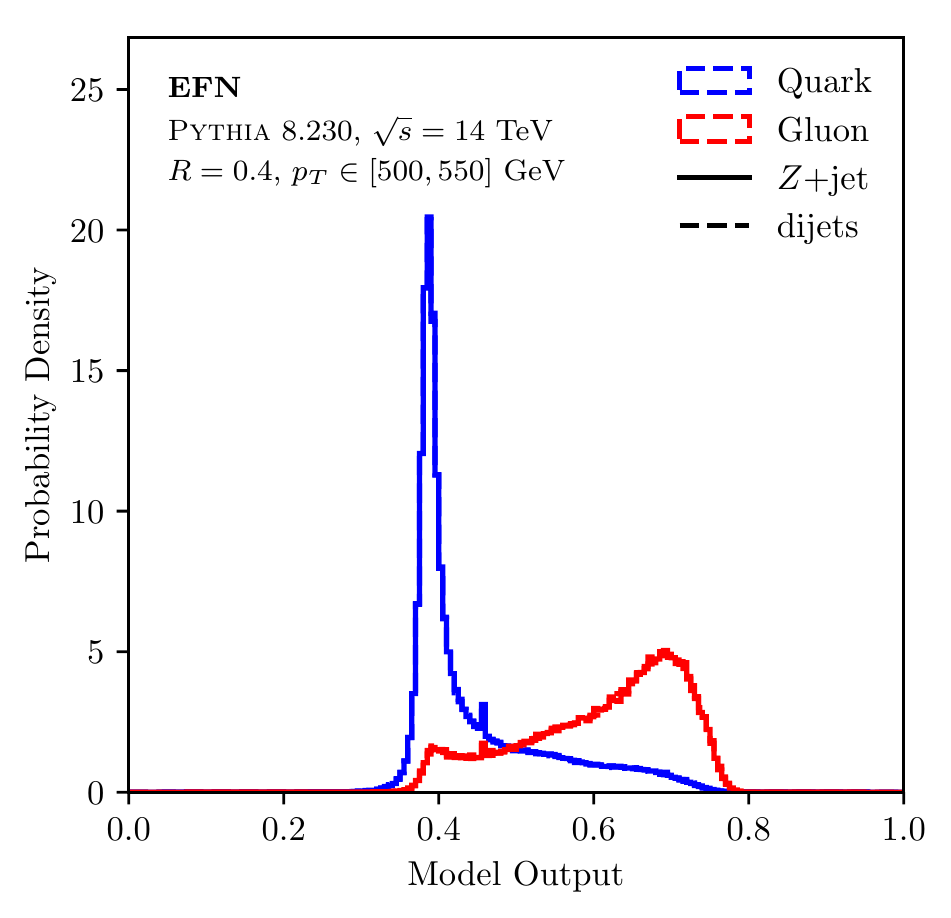}
\includegraphics[scale=.66]{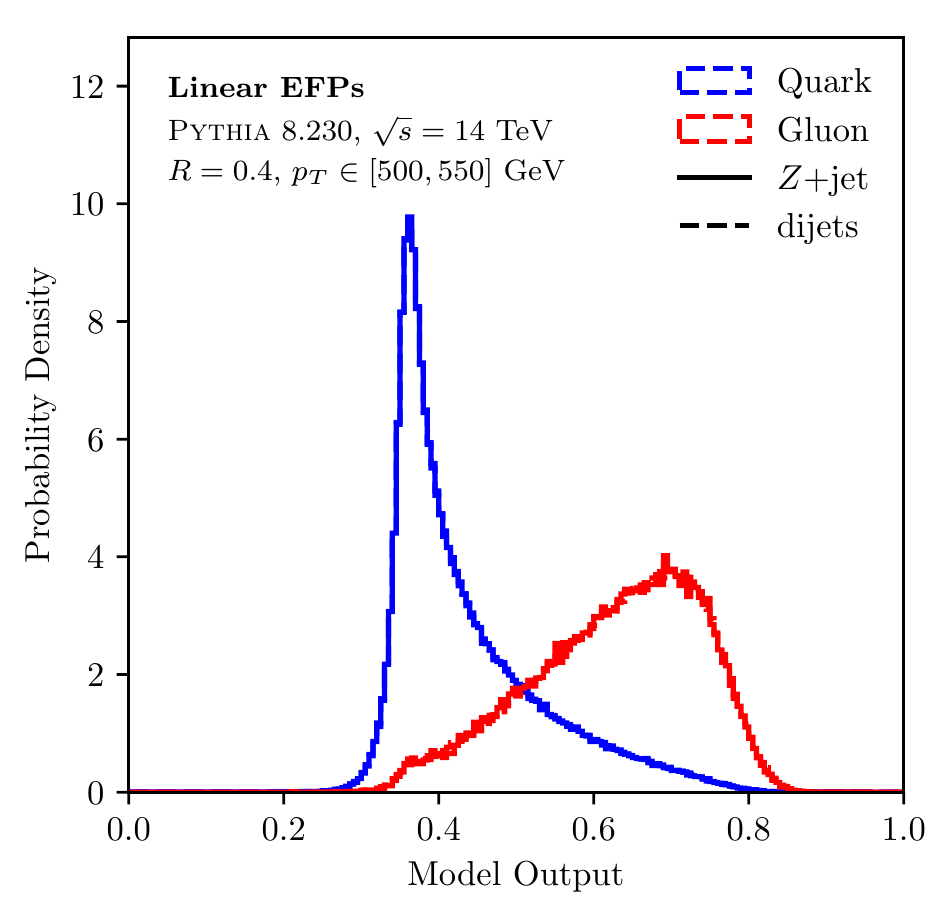}

\includegraphics[scale=.66]{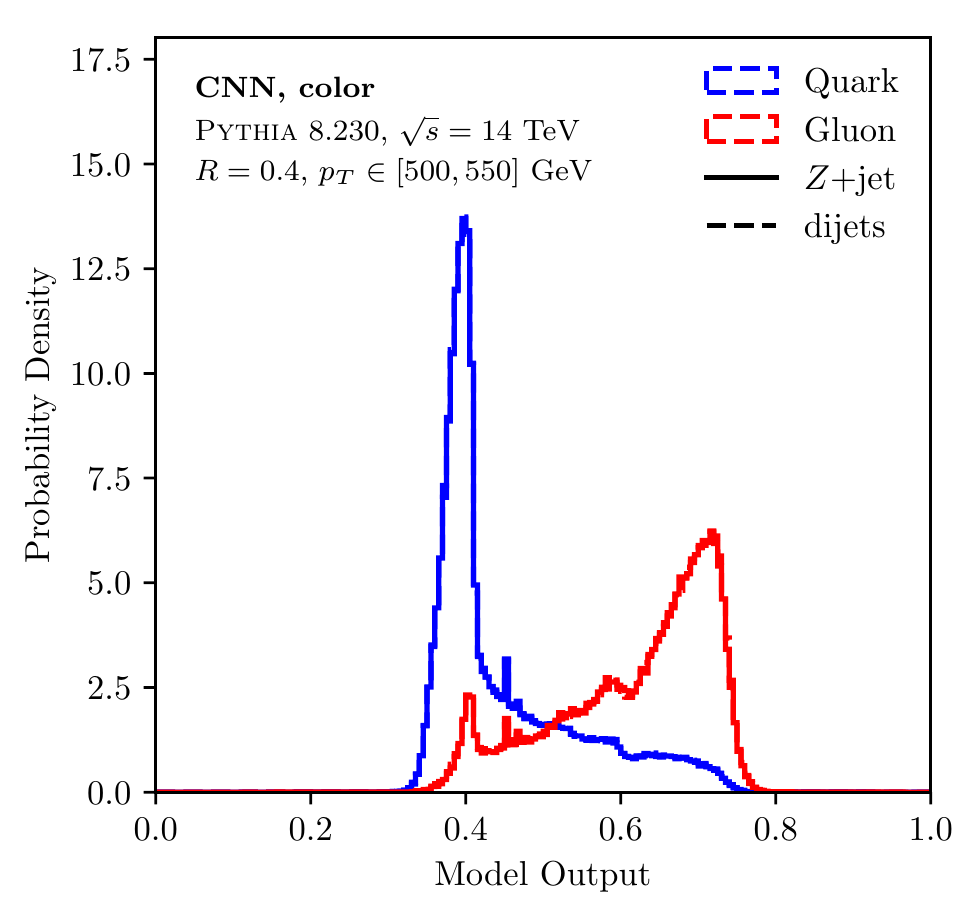}
\includegraphics[scale=.66]{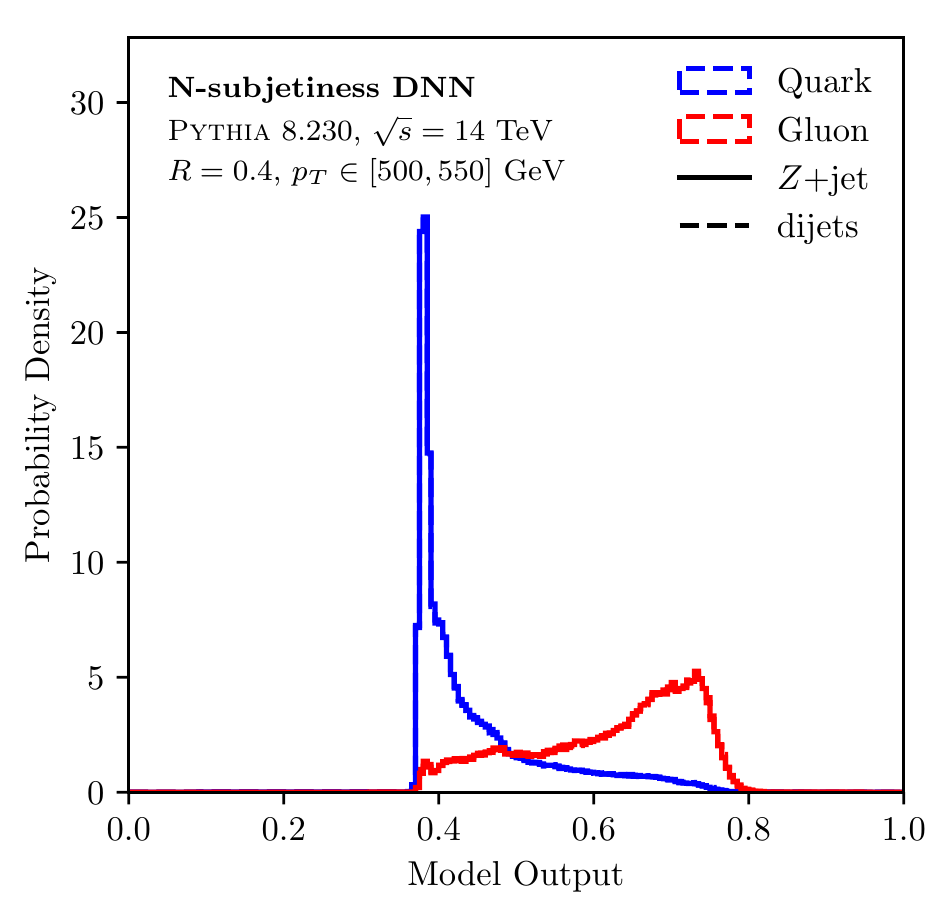}
\caption{Same as \Fig{fig:qgobshists} but for the six trained model outputs.}
\label{fig:qgmodelhists}
\end{figure}

In \Fig{fig:qgobshists}, we plot distributions for the six individual substructure observables, from both the $Z$+jet and dijet samples, showing the distributions separately for quarks and gluons as labeled by the \pythia hard scattering process.
Importantly, these distributions show a high degree of sample independence: the $Z$+jet and dijet quarks and gluons have very similar distributions.
In \Fig{fig:qgmodelhists}, we plot the distributions of the trained model outputs for quarks and gluons from both the dijet and $Z$+jet samples.
Similar to the standard jet observables in \Fig{fig:qgobshists}, a high degree of sample independence is observed.
This is perhaps more surprising than for the individual observables because these models have the ability to pick up on very slight differences as part of their training.
The observed amount of sample independence is encouraging for using CWoLa and jet topics with complicated models.

\begin{figure}[t]
\centering
\subfloat[]{\includegraphics[scale=.755]{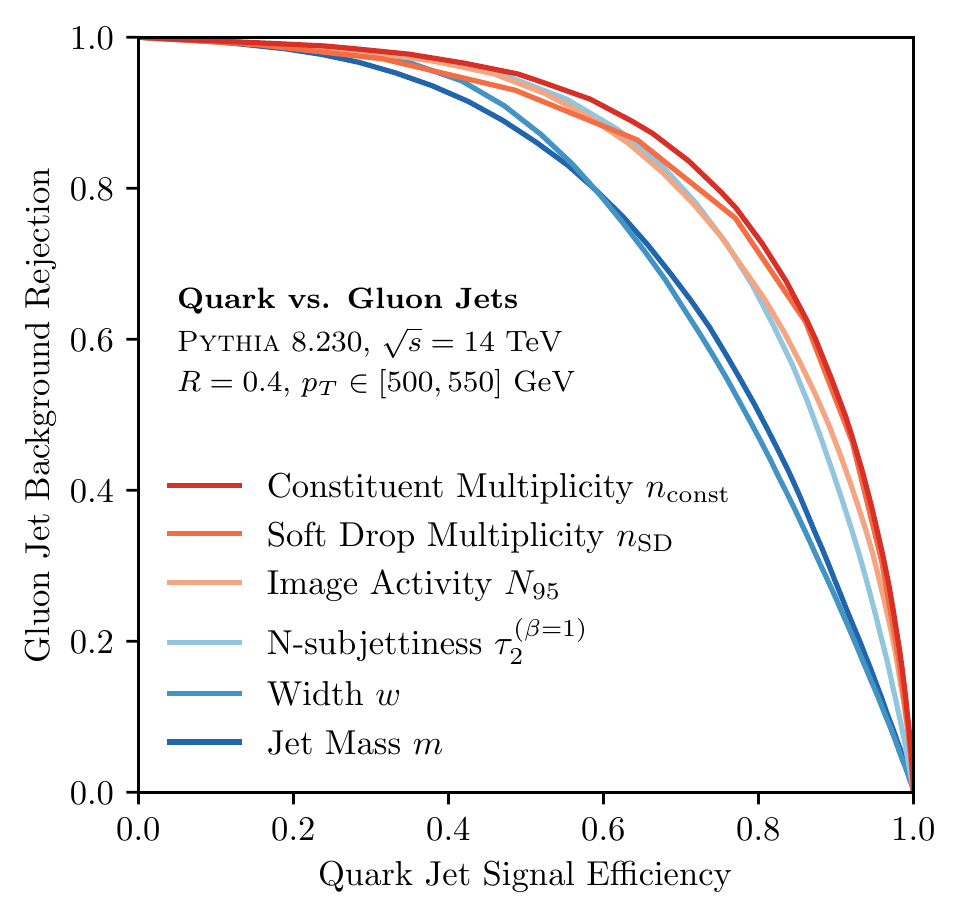}\label{fig:rocs-obs}}
\subfloat[]{\includegraphics[scale=.755]{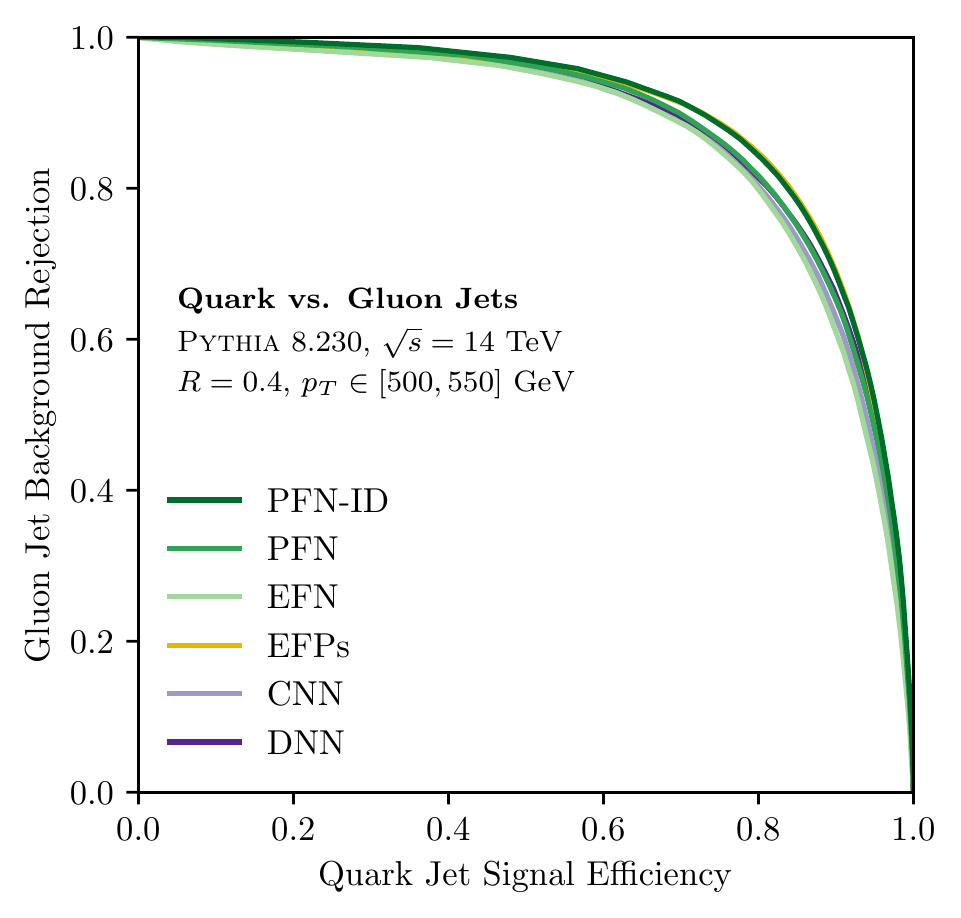}\label{fig:rocs-model}}
\caption{
ROC curves for (a) the individual jet observables and (b) the different models trained in the CWoLa paradigm to discriminate $Z$+jet and dijets events. 
These are calibrated using \pythia truth fractions.
The two best models are PFN-ID and EFPs, which are essentially on top of each other.
}
\label{fig:rocs}
\end{figure}

For completeness, we also show ROC curves for each of the observables and trained models in \Fig{fig:rocs}, calibrated using the \pythia fractions.
Specifically, we use the \pythia-labeled quark fractions of the $Z$+jet and dijet samples to calibrate the classifier ROC curve via \Eqs{eq:epb}{eq:eps}.
In \Fig{fig:rocs-obs}, we show ROC curves for each individual observable.
In \Fig{fig:rocs-model}, we show ROC curves for each of the trained models.

\clearpage

\bibliography{defineqg}

\end{document}